\documentclass[a4paper,twocolumn,prb,amssymb,groupedaddress,aps,superscriptaddress,nofootinbib,longbibliography,nobibnotes,noeprint]{revtex4-2}

\usepackage[utf8]{inputenc}
\usepackage{dcolumn}
\usepackage{bm}
\usepackage{graphicx,color}
\usepackage{soul}
\usepackage{epstopdf}
\usepackage{float}
\usepackage{placeins}
\usepackage{verbatim}
\usepackage{booktabs}
\usepackage[normalem]{ulem}
\usepackage[version=3]{mhchem}
\usepackage{amsmath}
\usepackage{multirow}
\usepackage{xcolor}

\newcommand{\cu}{Cu$_4$Cu}

\newcommand{\ergk}{ erg\,K/(mol\,G$^2$)}

\begin{document}

\title{Magnetism in a homometallic square-planar Cu$({\mathrm{II})}$$_4$Cu$({\mathrm{II})}$ metallacrown}

    \author{Luca Bischof}\email{Both authors contributed equally to this work.}
    \affiliation{Kirchhoff Institute for Physics, Heidelberg University, INF 227, 69120 Heidelberg, Germany}
    
    \author{Ariyan Tavakoli}\email{Both authors contributed equally to this work.}
    \affiliation{RPTU University Kaiserslautern-Landau, Department of Physics and Research Center OPTIMAS, 67663 Kaiserslautern, Germany}

	\author{Sascha Pietruska}
    \affiliation{Department of Chemistry, Johannes Gutenberg University of Mainz, 55128 Mainz, Germany}
    
    \author{Lena Spillecke}
    \affiliation{Kirchhoff Institute for Physics, Heidelberg University, INF 227, 69120 Heidelberg, Germany}
    
    \author{Ina Heckelmann}
    \affiliation{Kirchhoff Institute for Physics, Heidelberg University, INF 227, 69120 Heidelberg, Germany}
    
    \author{Changhyun Koo}
    \affiliation{Kirchhoff Institute for Physics, Heidelberg University, INF 227, 69120 Heidelberg, Germany}
    \affiliation{Department of Physics, POSTECH, Pohang 37673, Republic of Korea}
    
    \author{Luca M. Carrella}
    \affiliation{Department of Chemistry, Johannes Gutenberg University of Mainz, 55128 Mainz, Germany}
        
    \author{Eva Rentschler}\email{rentschl@uni-mainz.de}
    \affiliation{Department of Chemistry, Johannes Gutenberg University of Mainz, 55128 Mainz, Germany}
    
    \author{Hans Christian Schneider}\email{hc.schneider@rptu.de}
    \affiliation{RPTU University Kaiserslautern-Landau, Department of Physics and Research Center OPTIMAS, 67663 Kaiserslautern, Germany}
    
    \author{Rüdiger Klingeler}\email{klingeler@kip.uni-heidelberg.de}
    \affiliation{Kirchhoff Institute for Physics, Heidelberg University, INF 227, 69120 Heidelberg, Germany}

\begin{abstract}

The well-established homometallic metallacrown Cu(II)[12-MC\textsubscript{Cu(II)N(Shi)}-4]$^{2-}$, comprising four square-planarly arranged Cu(II) ions surrounding a slightly off-plane central Cu(II) ion, provides an excellent platform for investigating magnetism in exchange-coupled Cu(II) clusters both by experimental and \textit{ab initio} numerical methods. In the present work, sterically bulky substituents are introduced into the salicylhydroxamate ligand to obtain a modified derivative, Cu$({\mathrm{II})}$[12-MC\textsubscript{Cu$({\mathrm{II})}$N(L)}-4]$^{2-}$ (Cu$_4$Cu), thereby effectively suppressing intermolecular magnetic interactions and enabling a detailed investigation of its intrinsic ground-state spin structure and exchange interactions.  Static magnetization reveals competing antiferromagnetic exchange interactions of 
$J_1=-8.1(2)$~THz ($-33.5(8)$~meV) between the central and the peripheral sites
and $J_2=-4.40(12)$~THz ($-18.2(5)$~meV) between neighboring sites on the ring of the metallacrown, which are corroborated by \textit{ab initio} broken symmetry-based density functional theory (DFT). These couplings yield an $S=1/2$ ground state for Cu$_4$Cu at low temperatures, for which X-band and high-frequency/high-field electron paramagnetic resonance (EPR) precisely determine the $g$-factors $g_x = 2.028$, $g_y = 2.056$, and $g_z = 2.222$ close to axial anisotropy and in agreement with our DFT results. The ground-state spin-configuration is characterized by $d_{xy}$-character of the unoccupied states. Furthermore, our magnetization data evidence thermal transitions across a doublet-to-quartet gap of $\Delta_{1/2\rightarrow3/2} = 4.8(4)$~THz which is also indicated by the emergence of an excited-state EPR signal. 
\end{abstract}

\date{\today}
 \maketitle






\section{Introduction}

One of the most interesting classes of coordination compounds is that of metallacrowns, due to their structural and property versatility \cite{mezei_structural_2007}. These inorganic analogues of organic crown ethers consist of a repeating sequence, mostly [M–N–O–] or [M–N–N–], forming a macrocyclic ring. The ring size can vary from 9 to 60 atoms. The central cavity may be empty or accommodate transition-metal or lanthanide ions, resulting in a wide variety of structures.
Since the pioneering work of Pecoraro and Lah, interest in this class of compounds has steadily increased due to their potential applications \cite{lah_isolation_1989,Bodwin2001,Tegoni2012,Happ2015}. In recent years, the magnetic properties of metallacrowns have become a focus of research, as they can exhibit single-molecule magnetic behavior and could potentially act as qubits in quantum computing~\cite{zaleski_synthesis_2004,boron_mixed_2010,deb_understanding_2014,wang_opening_2021}. However, in multinuclear systems, understanding and precisely quantifying magnetic properties becomes increasingly challenging as the number of spins and centers increases. A less complex homometallic copper(II)-based metallacrown framework is the copper(II)-centered compound  Cu(II)[12-MC\textsubscript{Cu(II)N(Shi)}-4]$^{2-}$, which has already been studied extensively by some of us \cite{Happ2016,Ranecki2023,Puetz2025,Koo2019}.

\begin{figure*}[tb]
  \includegraphics[width=0.9\linewidth]{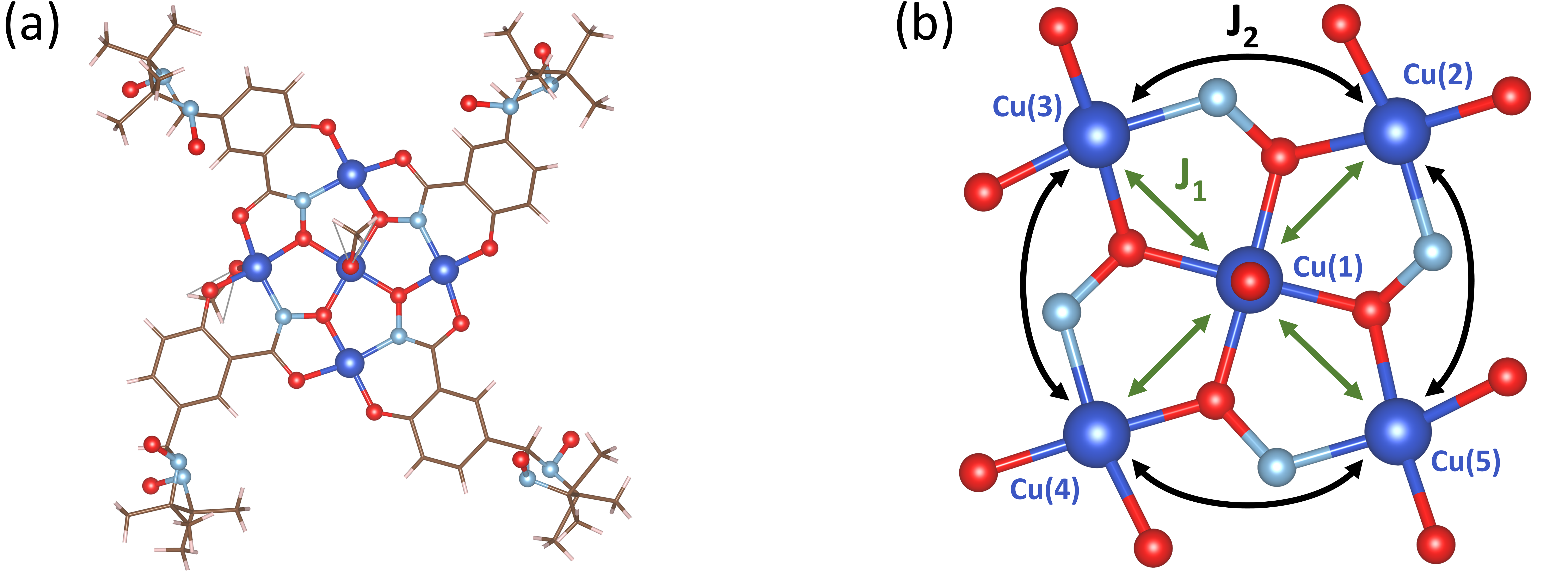}
  \caption{Top view of (a) the crystal structure and (b) the central molecular motif and magnetic couplings of \cu . In (a), a single complex is shown consisting of four repetition units [Cu–N–O] with ligand L. The crystal packing is shown in the SM~\cite{SM}. 
  The magnetic model in (b) is sketched with couplings as arrows indicating isotropic magnetic interactions between the central Cu$({\mathrm{II})}$ moment and its neighbors with coupling strength $J_1$ (green), and between neighboring outer Cu$({\mathrm{II})}$ moments with coupling strength $J_2$ (black). 
Ions and bonds coloring: Cu (blue), O (red), N (light blue), Si (green), C (brown), and H (white).}
  \label{Fig:structure}
\end{figure*}

In the present work, we investigate a derivative in which the ring-building salicylhydroxamate ligand carries bulky, silyl-protected tetramethylimidazolinyl side arms. The resulting compound Cu$({\mathrm{II})}$[12-MC\textsubscript{Cu$({\mathrm{II})}$N(L)}-4]$^{2-}$ (in short: \cu) contains four paramagnetic Cu(II) sites at the ring positions and a slightly off-plane central Cu(II) ion (see Fig.~\ref{Fig:structure}). The bulky substituents sterically shield the metallacrown cores, suppressing intermolecular magnetic interactions and enabling the magnetic properties of individual \cu\ metallacrowns to be investigated in the solid state. Employing X-band and high-frequency/high-field electron paramagnetic resonance (EPR) spectroscopy and static magnetometry as well as an \textit{ab initio}-based broken-symmetry approach, we resolve and quantify competing magnetic exchange interactions and $g$-factor anisotropy, thereby obtaining a comprehensive magnetic model for this five-spin system.

\section{Experimental methods}

All chemicals and solvents were purchased from Acros, Sigma Aldrich, Deutero, Alfa Aeser, Fluka, Fisher Scientific, Merck, Roth and TCI. Unless otherwise stated, reagents and solvents were used as received from the supplier. The ligand 5-(1,3-[(dimethyl-tert-butyl)silyl]oxy)-4,4,5,5-tetramethylimidazolinyl)\-salicylhydroxamic acid (L) was synthesized in five steps following reported procedures ~\cite{Sayre1955,Hirel2001,Ullman1972,Kolanji2017} (see the Supporting Information for details). Oxidative coupling of 2-nitropropane afforded 2,3-dimethyl-2,3-dinitrobutane, which was subsequently reduced with Zn/NH\textsubscript{4}Cl. The resulting crude 2,3-bis(hydroxylamino)-2,3-dimethylbutane was purified by Soxhlet extraction and condensed with methyl 5-formylsalicylate to yield methyl 5-(1,3-dihydroxy-4,4,5,5-tetramethylimidazolinyl)-salicylate through ring closure. Attempts to convert the ester directly into the hydroxamic acid were unsuccessful at both room and elevated temperatures under various reaction conditions. This may reflect decomposition of the imidazoline moiety due to the acidity of the hydroxamic acid \cite{Osiecki1968,Fidan2016}. The hydroxy groups were therefore protected by silylation. Subsequent treatment with hydroxylamine under basic conditions converted the ester into the hydroxamic acid, with concomitant cleavage of the phenolic silyl protecting group, affording ligand L.
The complex was synthesized following the general procedure previously reported by some of us \cite{Happ2014}. Dropwise addition of morpholine as a base to a solution of L and Cu(ClO\textsubscript{4})\textsubscript{2}·6H\textsubscript{2}O afforded the desired metallacrown  (C$_4$H$_{10}$NO)$_{2}$Cu$({\mathrm{II})}$[12-MC\textsubscript{Cu$({\mathrm{II})}$N(L)}-4] (\cu) as a green powder in good yield. Single crystals suitable for single-crystal X-ray diffraction were obtained by slow evaporation of the filtered solution.

Single crystal X-ray data for 
(C$_4$H$_{10}$NO)$_{2}$Cu$({\mathrm{II})}$[12-MC\textsubscript{Cu$({\mathrm{II})}$N(L)}-4]$\cdot$\textbf{solvent} (\cu$\cdot$\textbf{solvent}) were collected on a STOE IPDS 2T diffractometer (STOE \& Cie GmbH, Darmstadt, Germany) using a log-fine focus sealed X-ray tube with Mo–K$\alpha$ radiation ($\lambda = 0.71073$~\AA). Data collection was performed at 120(2)~K using an OXFORD Cryostream 700. Data integration, reduction and absorption correction was done with the X-Area 2.5.3.0 and LANA 2.7.5.0 software packages. 
The structure was solved with SHELXT~\cite{Sheldrick2015a} and refined by the full-matrix method based on F\textsuperscript{2} using SHELXL~\cite{Sheldrick2015b} implemented in the OLEX2\cite{Dolomanov2009} graphical interface. All non-hydrogen atoms were refined anisotropically, while the hydrogen atoms were fixed geometrically at calculated distances and allowed to ride on the parent atoms. Solvent masking was used to model the strongly disordered solvent molecules\cite{vanderSluis1990}. In total, 731 e\textsuperscript{-} per unit cell were counted, which correspond to four C\textsubscript{4}H\textsubscript{10}NO and 30 CH\textsubscript{4}O (736 e\textsuperscript{-}). Final crystallographic data are listed in the SM \cite{SM}. 
CCDC 2584802 contains the supplementary crystallographic data for this paper. This data can be obtained free of charge from The Cambridge Crystallographic Data Centre https://www.ccdc.cam.ac.uk/structures/.

Solid-state magnetic studies of \cu\ were performed on powdered and dried crystalline samples. Static magnetization data was obtained using a Quantum Design MPMS-XL SQUID magnetometer. Approximately 20~mg of the powdered samples was placed into a gelatin capsule, mounted in a plastic straw and fixed to the standard sample holder. The temperature and field dependent magnetic contribution of the holder was experimentally determined and subtracted from the measured susceptibility data. The experimental susceptibility data were corrected for the underlying diamagnetism using Pascal’s constants \cite{bain_diamagnetic_2008}. Variable temperature susceptibility data were collected in a temperature range of 2-300 K under an applied field of 0.1 T, while magnetization data were collected between 2 and 10 K and magnetic fields up to 7 T. 

X-band electron paramagnetic resonance (X-EPR) measurements were performed in an  Elexsys E500 spectrometer (Bruker) with an ER 4122SHQE CW cavity (quality factor $\approx 4000$) and at a frequency of $f = 9.6$~GHz. Low temperatures were achieved by a Cryo Edge cryostat equipped with a compressor system and the temperature
was controlled by means of a Mercury iTC (Oxford Instruments). The powder samples were fixed by n-eicosane (C\textsubscript{20}H\textsubscript{42}, CAS-Nr.112-95-8) in EPR quartz glass tubes (Sigma-Aldrich).

Continuous wave high-frequency/high-field electron paramagnetic resonance spectroscopy (HF-EPR) was performed using a millimeter vector network analyser (MVNA, ABmm) as phase-sensitive microwave source and detector~\cite{Comba_HFESR_2015}. Powder samples were prepared in a brass ring sealed by kapton tape, fixed with $n$-eicosane, and placed in the microwave path. Transmission spectra were obtained by simultaneously measuring the microwave amplitude and phase at several frequencies between 90~GHz and 330~GHz, and were corrected for background effects originating from the experimental setup. At each frequency, the spectra were recorded for both increasing and decreasing magnetic fields. Results obtained by fitting were averaged for both field sweep directions, while spectra for decreasing magnetic field are shown only. Temperature control at 2~K in magnetic fields up to 16 T was provided within a Variable Temperature Insert (VTI) of an Oxford magnet system equipped with a Nb\textsubscript{3}Sn superconducting coil~\cite{werner2017}.
Simulation and fitting of magnetization data and EPR spectra were performed using the MATLAB software package EasySpin \cite{stoll_easyspin_2006}. For simulating EPR spectra, an underlying Lorentzian isotropic broadening of $1$~mT (signal $s_1$) and $0.1$~mT (signal $s_2$) was employed to simulate the derivative spectra with dominant anisotropic broadening.

\section{Experimental results} \label{sec:Experimental results}

\subsection{Structure}


The core motif of the metallacrown compound 
(C$_4$H$_{10}$NO)$_{2}$Cu$({\mathrm{II})}$(MeOH)$_2$[12-MC\textsubscript{Cu$({\mathrm{II})}$N(L)}-4]$\cdot$\textbf{solvent} is very similar to those reported earlier by Pecoraro and some of us \cite{Happ2014,Koo2019}. The molecular scaffold consists of four  Cu$({\mathrm{II})}$ ions linked by the triple deprotonated ligand L, which forms the characteristic repetition unit [Cu-N-O]. An additional  Cu$({\mathrm{II})}$ ion occupies the resulting cavity (see Fig.~\ref{Fig:structure}). The negative charge of the complex Cu(MeOH)$_{2}$[12-MC$_{\rm Cu({\mathrm{II})}N(L)}$-4]$^{2-}$ is balanced by two morpholinium ions. The arrangement of the four ring copper ions creates a nearly perfect square with a planar structure. The distance between the copper atoms ranges from 4.563(1)~\AA\ to 4.577(1)~\AA. The central copper atom (II) is coordinated by four oxygen atoms from the hydroxamic acid groups and methanol molecules. Unlike the previously reported  Cu(II)[12-MC\textsubscript{Cu(II)N(Shi)}-4]$^{2-}$ the central copper ion lies 0.373(1)~\AA\ above the plane (see Fig.~S4 in \cite{SM}), while the  Cu$({\mathrm{II})}$–Cu$({\mathrm{II})}$ distances (peripheral to the inner  Cu$({\mathrm{II})}$) remain almost identical \cite{Happ2014}. We attribute this difference to the bulky substituent of the salicylic hydroxamic acid. The ligand backbone is slightly bent out of the plane, resulting in a slightly bowl-shaped motif for the complex. The 1,3-dimethyl-3-silylpropyl substituents are nearly perpendicular to the phenyl moieties. Thus, close contact between the metallacrowns above and below the plane is sterically hindered, which is crucial for precisely estimating the competing interactions within a metallacrown. The complex crystallizes in the triclinic space group P-1, where one entire metallacrown is part of the asymmetric unit. It should be noted that large solvent-accessible voids are present between the metallacrowns, accommodating two morpholinium counterions and 15 methanol molecules per asymmetric unit. The highly disordered solvent molecules and counterions could not be properly located and were refined using a solvent mask model. 


\subsection{Static Magnetic Properties} \label{sec: Static Magnetic Properties}

\begin{figure}[t]
	\centering
	\includegraphics[width=1\columnwidth]{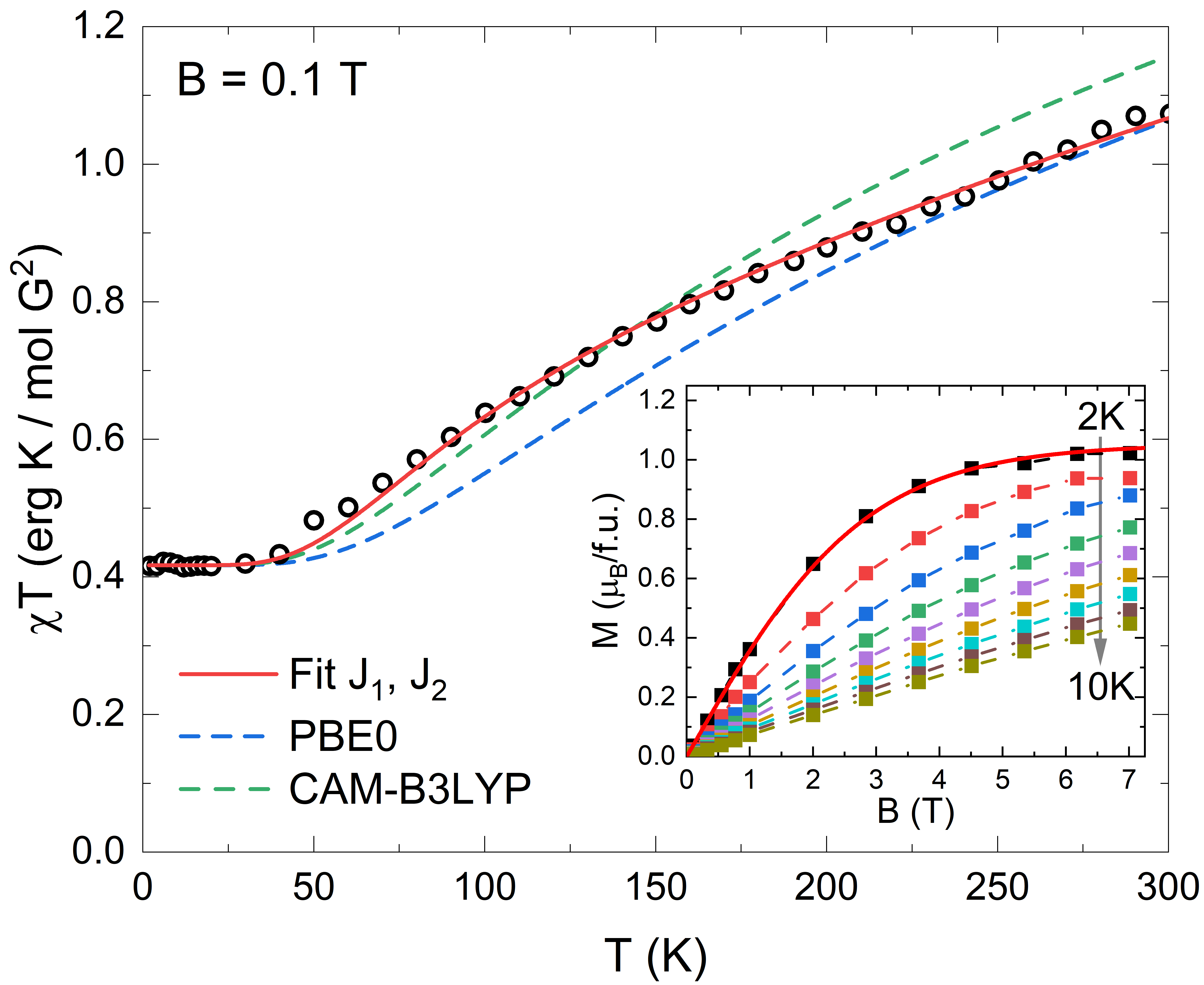}
	\caption{Temperature dependence of the static magnetic susceptibility measured at $B = 0.1$~T on a fixed powder sample is plotted as $\chi T$. The red solid line shows the best fit of the spin hamiltonian Eq.~\ref{eqn:Cu5_SH_SQUID} with the parameters described in the main text. Further lines display the simulated $\chi T$ curves with the magnetic exchange couplings $J_1,J_2$ calculated by PBE0 (blue) and CAM-B3LYP (green) as described in Sec.~\ref{ch:numerics}. Inset: Isothermal magnetization at low temperatures from 2~K to 10~K with a temperature increment of 1~K between the curves. The red solid line shows the  fit of the Brillouin function Eq.~\ref{eqn: Brillouin} to the data at $T=2$~K.
    }
	\label{Fig:Cu5_dc_susceptibility}
\end{figure}

The temperature dependence of the static magnetic susceptibility $\chi=M/B$ was
measured in an external magnetic field of $B = 0.1$~T on a fixed powder sample (see Fig.~\ref{Fig:Cu5_dc_susceptibility}). Up to 30~K, the $\chi$T product is temperature independent at $\chi T=0.417(5)$~\ergk, which suggests an $S = 1/2$ ground state with $g = 2.109(13)$ derived from Curie behavior $\chi T = C = N_\text{A}g^2\mu_\text{B}^2S(S+1) /3k_\text{B}$. Upon further heating, $\chi T$ rises and develops a broad hump centred around 100 to 150~K. 
While this behaviour indicates thermal excitations of the antiferromagnetically coupled ground state, the absence of saturation in $\chi T$ even at room temperature implies strong antiferromagnetic coupling between the Cu$({\mathrm{II})}$ spins. 

For molecules with the spin density localized around specific sites, the magnetic properties can often be described by an effective Heisenberg model defined in the following spin Hamiltonian: 
\begin{equation}
	\mathcal{H} = g \mu_{\mathrm{B}} \sum_i \mathbf{S}_i \cdot \mathbf{B}-\sum_{i<j} 
		 J_{ij} \,\mathbf{S}_i \cdot \mathbf{S}_j.
	\label{eqn:Cu5_SH_SQUID}
\end{equation}
Here, the first term is the Zeeman interaction of the external magnetic field $\mathbf{B}$ with the Cu(II) spins $\mathbf{S}_i$ with spin number $1/2$. The second term describes the magnetic exchange interactions between spin centers $i$ and $j$. In \cu, these interactions can be modeled by two exchange parameters, as illustrated in Fig.~\ref{Fig:structure}(b): the couplings between the central and peripheral Cu(${\mathrm{II}}$) ions  ($J_1$) and the interactions between neighboring Cu(${\mathrm{II}}$) ions within the outer ring ($J_2$), respectively, assuming an isotropic spin system. 

The exchange coupling constants $J_1$ and $J_2$ can be obtained from a fit of the Heisenberg model \eqref{eqn:Cu5_SH_SQUID} to the static magnetization in Fig.~\ref{Fig:Cu5_dc_susceptibility}. To reduce various parameter dependencies, an isotropic $g$-factor is assumed for all five  Cu$({\mathrm{II})}$ and fixed to the effective value $g = 2.109(13)$ obtained from the low-temperature $\chi T$ data.
The fitting (red line in Fig.~\ref{Fig:Cu5_dc_susceptibility}) yields  
$J_1=-8.1(2)$~THz ($-33.5(8)$~meV) 
and 
$J_2=-4.40(12)$~THz ($-18.2(5)$~meV). 
These values as well as the ratio $J_1$/$J_2$ = 1.84(7) are in the range observed for antiferromagnetically coupled, square planar  Cu$({\mathrm{II})}$ based metallacrown complexes reported before~\cite{Koo2019,song_preparation_2000,pavlishchuk_magnetic_2011,mcdonald_progressive_2013,liu_crystal_2020}. Note that fitting with the $g$-factor as a free parameter leads to coinciding interaction parameters within the error bars. 

The inset in Fig.~\ref{Fig:Cu5_dc_susceptibility} depicts the isothermal magnetization up to 7~T and at 2~$\text{K}\leq T \leq 10~\text{K}$. At the lowest temperature of $T = 2$~K, the magnetization curve flattens into saturation behavior at high field with $M(B=7\text{~T}) = 1.02~\mu_\text{B}\text{/f.u.}$. This value confirms the presence of a spin $S = 1/2$ ground state in accordance with the low-temperature $\chi T$($T$) data. The isothermal magnetization can be quantitatively described in terms of the Brillouin-function $B_S(x)$ for $S=1/2$:
\begin{equation}
	M(B) = M_{\text{s}} \cdot B_S\left(\frac{g\mu_{\mathrm{B}}SB}{k_{\mathrm{B}}T}\right)
	\label{eqn: Brillouin}
\end{equation}
with the saturation magnetization $M_s$, the Bohr magneton $\mu_{\rm B}$ and the Boltzmann constant $k_{\rm B}$. The excellent agreement of this response function of free spins with our data further corroborates the $S=1/2$ ground state at low temperatures and confirms the absence of intermolecular magnetic interaction -- as intended when using bulky ligands for \cu. From fitting with \eqref{eqn: Brillouin} at $T=2$~K, we obtain the g-factor $g_{M(B)} = 2.10(2)$, which is in perfect agreement with the value obtained from the $\chi T$ data.

\subsection{EPR Spectroscopy} \label{sec:EPR-Spectroscopy}

\begin{figure}[bt]
	\centering
	\includegraphics[width=1\columnwidth]{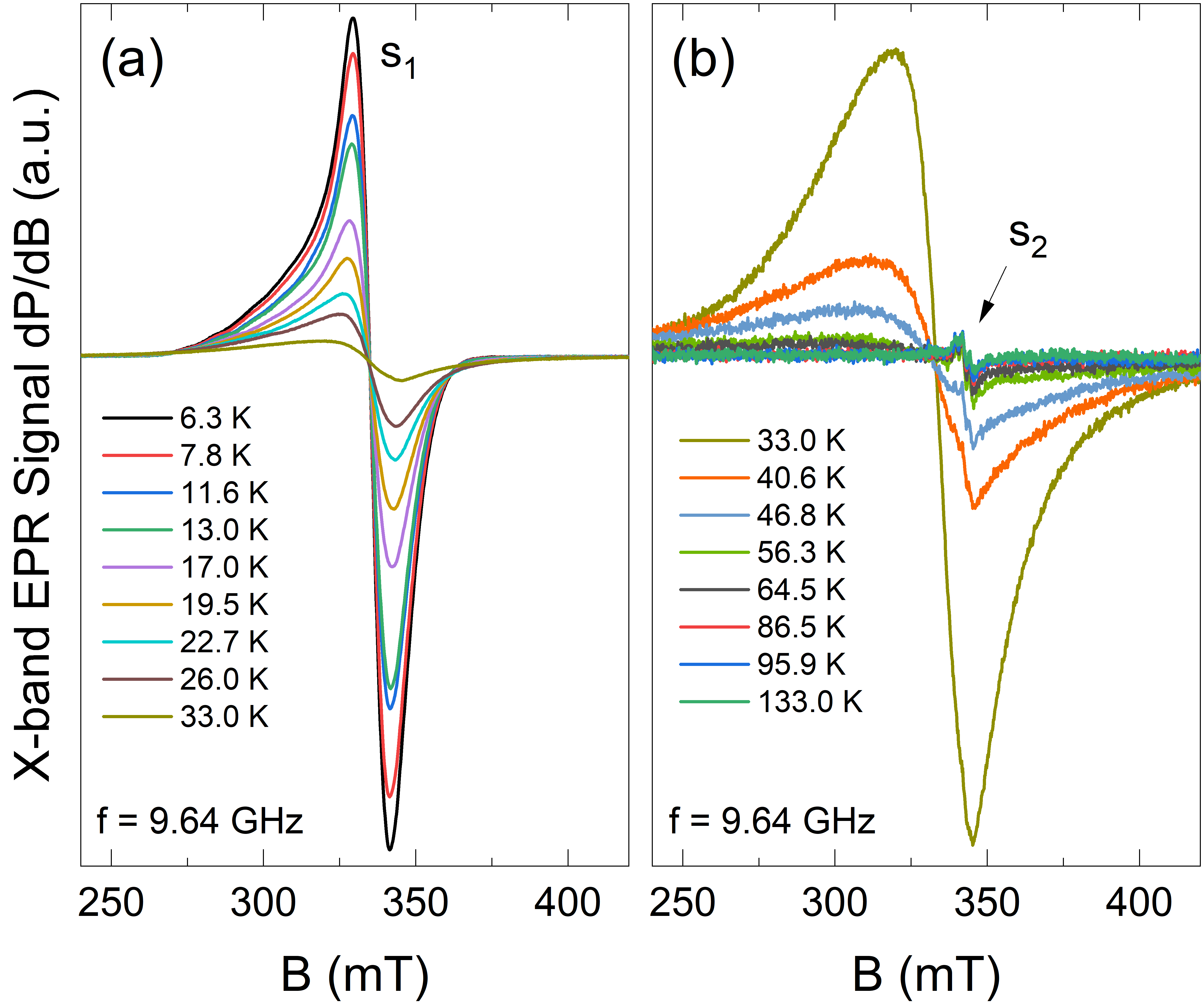}
	\caption{X-band EPR spectra of fixed powder sample of Cu$^{\mathrm{II}}$[12-MC$_{Cu^{\mathrm{II}}}$-4] at (a) low and (b) intermediate temperatures exhibit the strong low-temperature resonance line $s_1$ and the thermally-induced feature $s_2$. Note different ordinates in (a) and (b) such that the spectrum at 33~K is the least intense spectrum in (a) and the most intense spectrum in (b).    
    }
	\label{Fig:Cu5_EPR_all_spectra}
\end{figure}

X-band EPR measurements (Fig.~\ref{Fig:Cu5_EPR_all_spectra})  were performed at different temperatures to investigate the magnetic properties of \cu\ in more detail. In Fig.~\ref{Fig:Cu5_EPR_all_spectra}(a), one distinct resonance signal ($s_1$) is observed at low temperatures. The signal is anisotropic as it exhibits enhanced spectral weight at the high magnetic field side compared to the low-field side. With increasing temperature, the signal intensity strongly decreases, as also visible in Fig.~\ref{Fig:Cu5_EPR_all_spectra}(b) where the spectra at $T\geq 33$~K are presented with appropriate scaling of the ordinate. A second resonance feature ($s_2$) is observed for increasing temperature as indicated by the black arrow in Fig.~\ref{Fig:Cu5_EPR_all_spectra}(b) and superposes the strong low-temperature signal $s_1$. This thermally-induced feature $s_2$ appears between 33~K and 40~K and exhibits a fine structure discussed below.

\begin{figure}[bt]
	\centering
	\includegraphics[width=1\columnwidth]{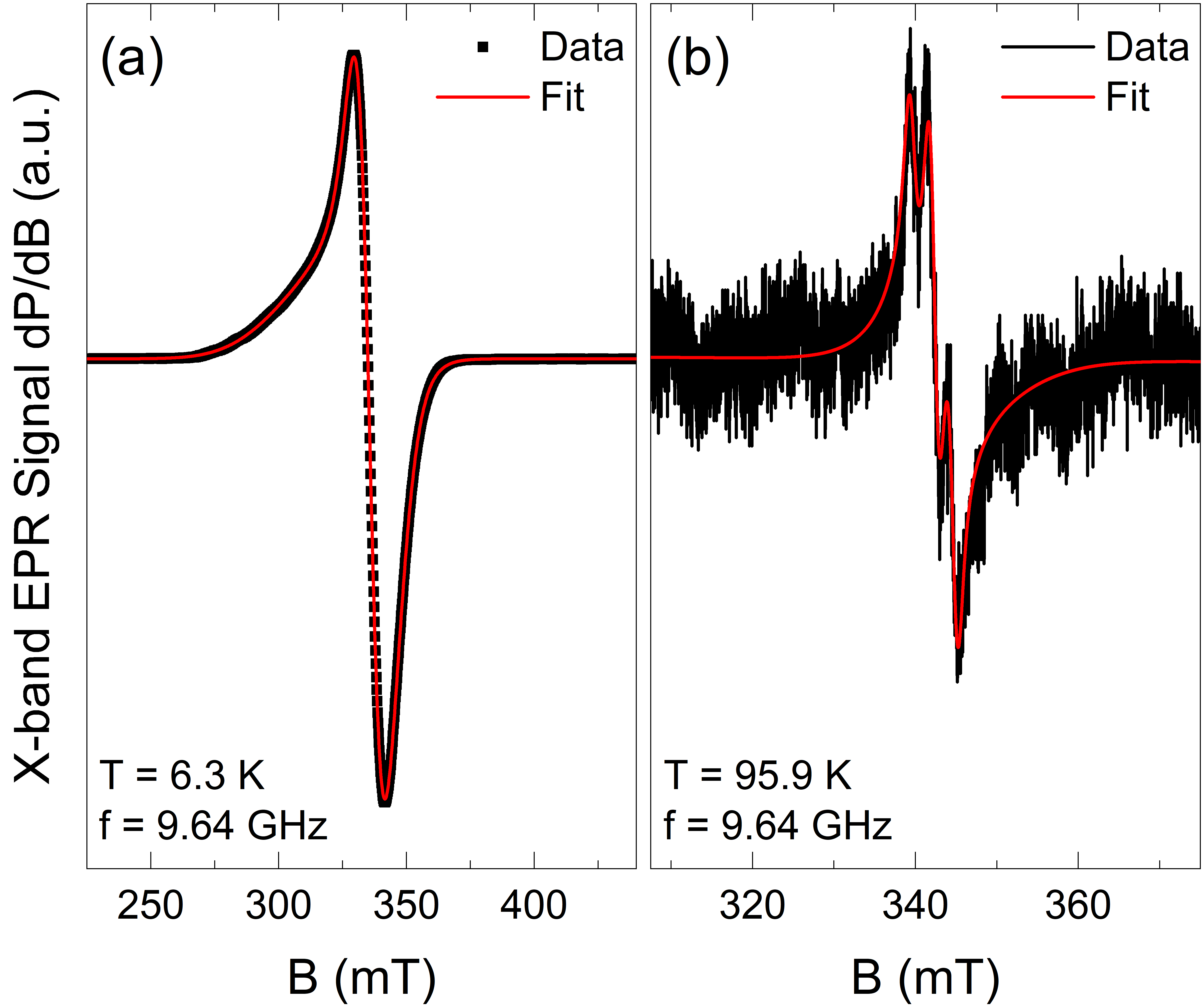}
	\caption{Fitting of the X-band EPR spectra showing (a) the low-temperature signal $s_1$ at $T = 6.3$~K and (b) the thermally-induced resonance line $s_2$ at $T = 95.9$~K. Solid red lines show the best fits to the data based on an $S = 1/2$ spin model with anisotropic g-factors, line broadening and hyperfine interaction (only b) as described in the main text. The resulting spin Hamiltonian parameters are summarized in Tab.~\ref{tab:Cu5_fitparameter}. }
	\label{Fig:Cu5_EPR_X-EPR_fitting}
\end{figure}

The observed EPR features $s_1$ and $s_2$ reveal detailed information on the spin system of \cu. The low-temperature resonance line $s_1$ at $T=6.3$~K can be excellently described by an $S=1/2$ spin hamiltonian with anisotropic $g$-factors and anisotropic line broadening. We attribute the latter to non-vanishing hyperfine interaction associated with the finite nuclear spin of the  Cu$({\mathrm{II})}$ ions. The best fit to the spectrum is shown by the red line in Fig.~\ref{Fig:Cu5_EPR_X-EPR_fitting}(a) with the obtained parameters summarized in Tab.~\ref{tab:Cu5_fitparameter}. The results  confirm the $S=1/2$ ground state. Furthermore, the obtained $g$-factors $g_x = 2.028$, $g_y = 2.056$, and $g_z = 2.222$ are close to axial anisotropy. The line broadening has a dominant component in $z$ direction but smaller $x$ and $y$ components suggesting that the hyperfine interaction strength figures the same anisotropy. The $g$-factors from EPR correspond to an effective $g$-factor  $g_\text{eff}=2.103$ which agrees with the value derived from the analysis of the static magnetization and is typical for Cu(II) in distorted octahedral or square-pyramidal environment~\cite{Ju2024,koo2016,Wang2018}.
They further are close to the $g$-factors  $g_x=2.03(1), g_y=2.04(1), g_z=2.23(1)$ reported for the   Cu(II)[12-MC\textsubscript{Cu(II)N(Shi)}-4]$^{2-}$ metallacrown complex.~\cite{Koo2019}

\begin{table}[b]
	\centering
	\caption[]{Spin Hamiltonian parameters for the low-temperature EPR signals $s_1$ and the thermally-induced feature $s_2$, respectively: Anisotropic $g$-factors, anisotropic line broadening parameters $H_{\mathrm{strain}}$, and the hyperfine coupling tensor $A$ are from best fits to the spectra (see Fig.~\ref{Fig:Cu5_EPR_X-EPR_fitting}) as described in the main text.}

		\begin{tabular}{c|c|ccc}   \hline\hline     
		\;\;signal\;\;& & $x$ & $y$ & $z$ \\			\hline
		  \multirow{2}{*}{$s_1$}&$g$ & \;\;2.028\;\; & \;\;2.056\;\; & \;\;2.222\;\; \\
		  &$H_{\mathrm{strain}}$ (MHz) & 432& 250 & 1411 \\       \hline
          
		  &$g$ & 2.010 & 2.013 & 1.985\\
		  $s_2$ &\quad$H_{\mathrm{strain}}$ (MHz)\quad & 38 & 29 & 524\\
        &$A$ (MHz) & 23 & 74 & 96  \\
		  
		\hline\hline
	\end{tabular}

	\label{tab:Cu5_fitparameter}
\end{table}

\begin{figure}[bt]
	\centering
	\includegraphics[width=1\columnwidth]{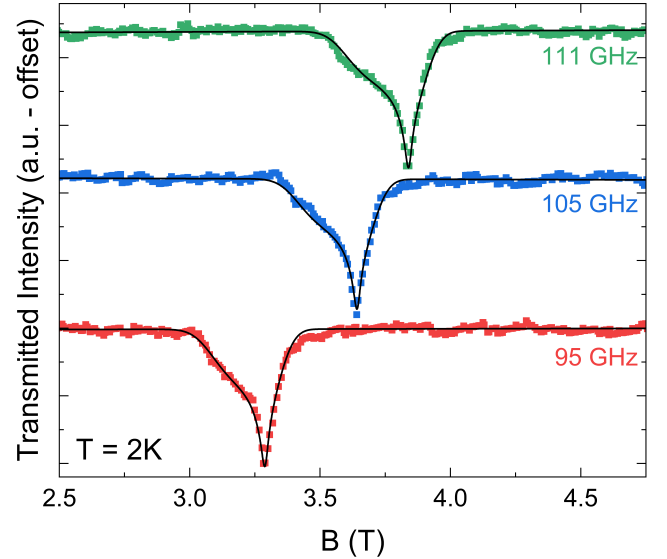}
	\caption{High-frequency/high-field EPR spectra of \cu\ at $T=2$~K exhibit the low-temperature EPR signal $s_1$. Solid black lines describe the best simultaneous fit to the three transmission spectra with microwave frequencies around 100~GHz, which yields the $g$-factors discussed in the main text. }
	\label{Fig:Cu5_HFEPR}
\end{figure}

The low-temperature signal $s_1$ observed in X-band EPR was further detected in HF-EPR measurements for frequencies between 90~GHz and 330~GHz. Best signal-to-noise ratios were found in three spectra at frequencies around 100~GHz (see Fig.~\ref{Fig:Cu5_HFEPR}), which were simultaneously fitted by the $S=1/2$ model used for the X-EPR data of $s_1$. Note that HF-EPR spectra shown in Fig.~\ref{Fig:Cu5_HFEPR} display microwave transmission, while X-EPR results in derivative absorption spectra. The resulting $g$-factors $g_x^\text{HF} = 2.013$, $g_y^\text{HF} = 2.051$, and $g_z^\text{HF} = 2.188$ are in agreement to the ones obtained from X-EPR, see Tab.~\ref{tab:Cu5_fitparameter}. Together, X-band EPR and HF-EPR thus unanimously confirm the $S=1/2$ ground state of \cu\ at low temperatures and allow the precise determination of the ground state's $g$-factors.

The thermally-induced signal $s_2$ is best discussed in detail at $T = 95.9$~K (see Fig.~\ref{Fig:Cu5_EPR_X-EPR_fitting}(b)), as the intensity of the superimposed low-temperature resonance $s_1$ decreases with temperature and almost completely vanishes towards $100$~K. The resonance line $s_2$ is centered around $342$~mT ($g=2.01$) and is dominated by three sharp maxima and three sharp minima, which come along with additional broadening at the high-field side. These features point towards hyperfine interaction with a nuclear spin of $I=1$ as present in $^{14}$N. Indeed, the spectra can be well described by a spin $S=1/2$ model with anisotropic hyperfine coupling to one $I=1$ nucleus, anisotropic $g$-factors and anisotropic line broadening (see Fig.~\ref{Fig:Cu5_EPR_X-EPR_fitting}(b)). The Hamiltonian parameters obtained from the best fit are summarized in Tab.~\ref{tab:Cu5_fitparameter}. As for $s_1$, the $g$-factors are close to axial anisotropy ($g_x = 2.010$, $g_y = 2.013$, $g_z = 1.985$) but with the distinction that the out-of-plane $g$-factor $g_z$ is smaller than the in-plane components. Small in-plane line broadening enables the resolution of the hyperfine structure, while the out-of-plane line broadening is -- as for $s_1$ -- distinctly larger and corresponds to the broad features of the spectrum.
Despite the satisfactory fitting of the spectrum, it remains elusive why coupling to one single $I=1$ nucleus is dominant, while four N ($I=1$) and five Cu ($I=3/2$) are present in the ring and cavity of \cu\ (see Fig.~\ref{Fig:structure}). One might speculate that additional, weaker spectral features are present but obscured by the noise.

\section{Ab initio results}\label{ch:numerics}

\subsection{Electronic Ground State and Magnetic Exchange Couplings\label{sec:ground-state}}

In this section, we relate the magnetic properties to \textit{ab initio} calculations of the electronic structure of the molecules. 
In particular, we compute the exchange coupling parameters contained in the Heisenberg Hamiltonian Eq.~\eqref{eqn:Cu5_SH_SQUID} from DFT calculations for broken-symmetry configurations for the spins 
of the Cu(${\mathrm{II}}$) ions. We follow the general approach of Ref.~\cite{Pavlyukh2019} , which is based on the analysis of specific spin dimers and relating the energy differences between electronic states obtained through the broken symmetry method with the corresponding Heisenberg exchange parameters~\cite{Noodleman1981,Noodleman1986,Dai2003,WHANGBO2003}. 

Within the broken-symmetry (BS) method, the spin system is divided into effective spin dimers denoted by $A$ and $B$ to define subsets of interacting magnetic centers. Based on this classification, different spin configurations can be introduced. In particular, a high-spin (HS) state with total spin $S = S_A + S_B$, and 
BS states 
in which opposite spin orientations are enforced on the selected subsets of spins.
Importantly, the BS states are not eigenstates of the Heisenberg Hamiltonian nor of $S^2$. The energies of the HS and BS electronic states are evaluated using density functional theory (DFT) and these results are used for the determination of the exchange parameters as follows.  

According to the mapping analysis presented by Dai et al.~\cite{Dai2003}, the expectation value for a broken symmetry spin state for the spin dimer with $M$ spins at spin site $A$ and $N$ spins at spin site $B$ can be modeled as a single determinant:
\begin{equation}
	 \langle S^2 \rangle_{\text{BS}} = \frac{(M - N)^2}{4} + \frac{M + N}{2}.
	\label{eqn:S2_value}
\end{equation}
The final exchange integral $J_{AB}$ between the two spin centers $A$ and $B$ is then derived through  so-called "Yamaguchi" equation~\cite{Yamaguchi2018}:   
\begin{equation} \label{eqn:J_AB}
J_{AB} =
\frac{E_{\mathrm{HS}} - E_{\mathrm{BS}}}
{\langle S^2 \rangle_{\mathrm{HS}} - \langle S^2 \rangle_{\mathrm{BS}}}.
\end{equation}

Here, $E_{\mathrm{HS}}$ and $E_{\mathrm{BS}}$ denote the total energies of the high-spin (ferromagnetic) state and the broken-symmetry state, respectively.
\begin{figure}[t]
	\centering
	\includegraphics[width=1\columnwidth]{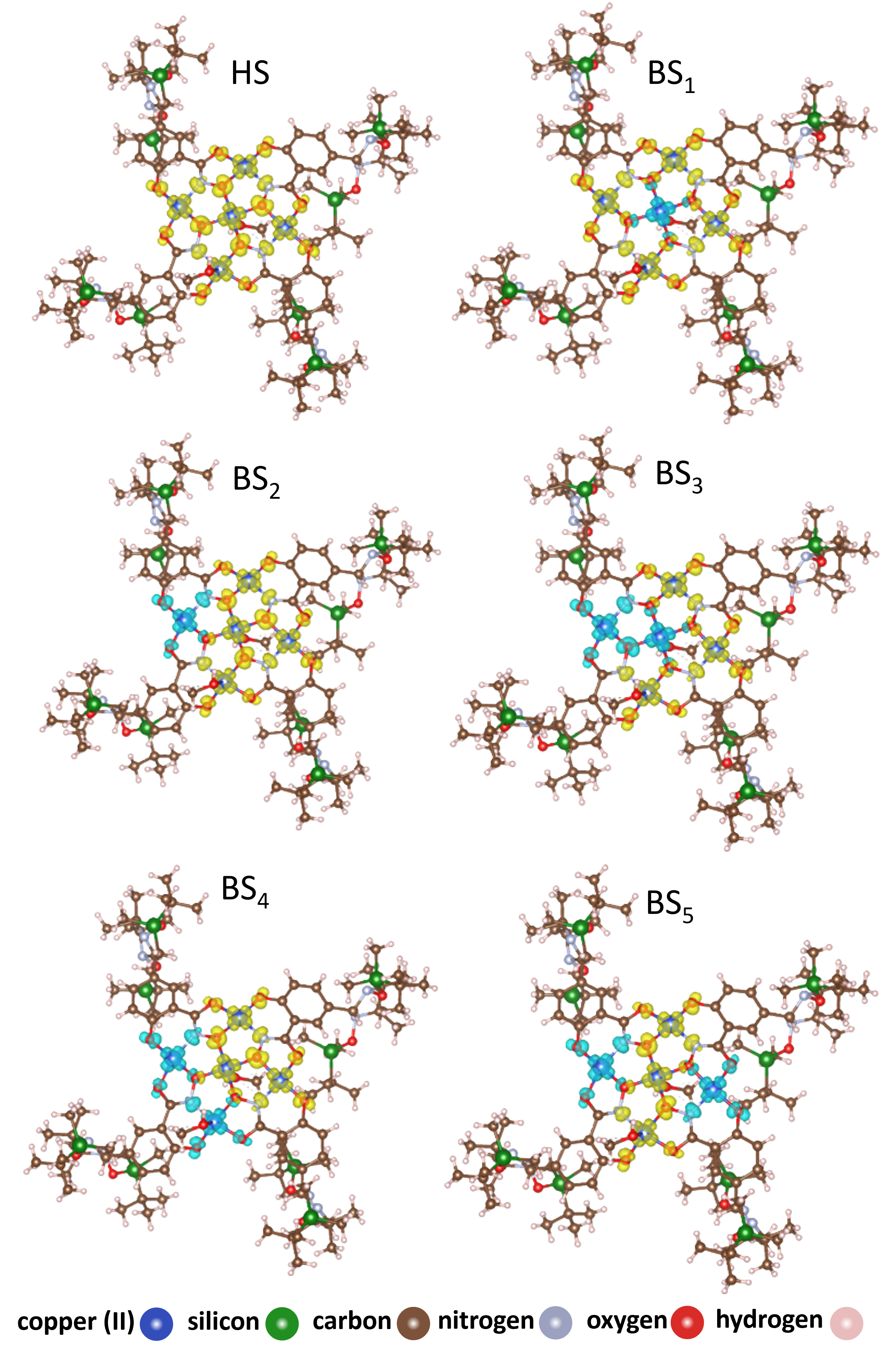}
	\caption{Spin density isosurfaces for all different spin configurations presented in Table~\ref{tab:relative_energies}. Yellow and light blue surfaces represent spin-up and spin-down density, respectively. Denoting the central and peripheral Cu ions as A and B, respectively,
    one observes a spin localization along the A--O--B and A--O--N--B bridges as would be expected for a superexchange coupling mechanism~\cite{goodenough,Pavlyukh2019}. See Sec.~\ref{sec:electronic structure} for more details on electronic structure. }
	\label{Fig:spin_densities}
\end{figure}
For the five site system, the $\langle S^2 \rangle$ expectation values for each corresponding spin states are obtained by substituting the appropriate $M$ and $N$ values into Eq.~\ref{eqn:S2_value}, e.g., the HS state ($M=5, N=0$), configurations with one flipped spin ($M=4, N=1$) and configurations with two flipped spins ($M=3, N=2$). The calculated $\langle S^2 \rangle$ expectation values for each spin states are  reported in Table~\ref{tab:relative_energies}.

In the following, our aim is to determine the total energies of the HS (ferromagnetic) and BS states, appearing in the numerator of Eq.~\ref{eqn:J_AB} to calculate not only the exchange coupling constants but also to identify the ground state energy. As in Refs.~\cite{Tavakoli, Pavlyukh2019}, six distinct spin configurations were considered for the five Cu(${\mathrm{II}}$) ions centers in the \cu\ molecule. These include one high-spin (HS) state and five broken-symmetry (BS) states, which are listed in Table~\ref{tab:relative_energies}. The state with the lowest energy for the molecular structure shown in Fig.~\ref{Fig:structure} is the BS5 configuration. As stressed by Perdew et al.,~\cite{Perdew_2021}, these near-eigenstates of broken symmetry that result from a density functional calculation are a sign of strong correlations among the electrons in a symmetry-unbroken wavefunction. 
\begin{figure}[t]
	\centering
	\includegraphics[width=1\columnwidth]{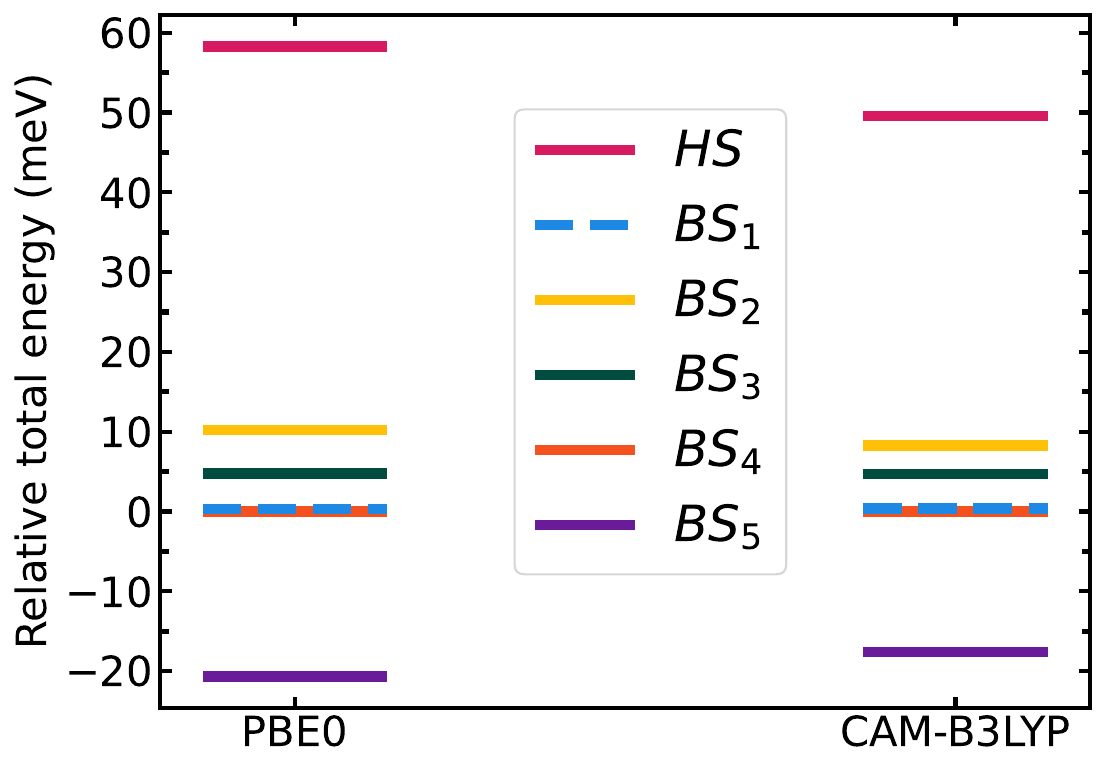}
	\caption{ 
    Calculated total energies relative to the BS$_4$ spin configuration of using PBE0 and CAM-B3LYP exchange functionals (see also Table~\ref{tab:relative_energies}). 
    }
	
    \label{Fig:Broken_sym_energies}
\end{figure}
The calculation of the energies of HS and BS states is carried out within the framework of DFT using the \textsc{ORCA} quantum chemistry program~\cite{ORCA}. This approach allows for the use of hybrid exchange--correlation functionals together with relativistic corrections within an all-electron framework. In particular, the hybrid functionals PBE0~\cite{PBE0} and the range-separated CAM-B3LYP~\cite{CAM-B3LYP} were used. Relativistic effects were taken also into account using the zero-order regular approximation. To this end, the ZORA-def2-SVP basis set was adopted, which incorporates scalar relativistic effects and is suitable for describing spin--orbit coupling contributions, especially in systems containing heavier elements. 

\begin{table}[b]
\centering
\caption{ Calculated relative total energies for different spin configurations, obtained using PBE0 and CAM-B3LYP hybrid functionals. All energies are given relative to the BS$_4$ state. The corresponding $\langle S^2 \rangle$ expectation values for each spin configuration are also shown. }
\label{tab:relative_energies}
\begin{tabular}{l|c|c|c}
\hline\hline
States& $\langle S^2 \rangle$ & $E_{\rm PBE0}$ (meV)  & $E_{\rm CAM-B3LYP}$ (meV) \\
\hline
HS $ =  \begin{pmatrix} \uparrow & & \uparrow \\ & \uparrow & \\ \uparrow & & \uparrow \end{pmatrix}$ & $\frac{35}{4}$ &  58.27  &  49.52 \\
\hline
BS$_1 = \begin{pmatrix} \uparrow & & \uparrow \\ & \downarrow & \\ \uparrow & & \uparrow \end{pmatrix}$ & $\frac{19}{4}$ &  0.28  &   0.39 \\
\hline
BS$_2 = \begin{pmatrix} \downarrow & & \uparrow \\ & \uparrow & \\ \uparrow & & \uparrow \end{pmatrix}$ & $\frac{19}{4}$ &  10.18   &   8.29 \\
\hline
BS$_3 = \begin{pmatrix} \downarrow & & \uparrow \\ & \downarrow & \\ \uparrow & & \uparrow \end{pmatrix}$ & $\frac{11}{4}$ &  4.77   &   4.71 \\
\hline
BS$_4 = \begin{pmatrix} \downarrow & & \downarrow \\ & \uparrow & \\ \uparrow & & \uparrow \end{pmatrix}$ & $\frac{11}{4}$ &  0.00   &   0.00 \\
\hline
BS$_5 = \begin{pmatrix} \uparrow & & \downarrow \\ & \uparrow & \\ \downarrow & & \uparrow \end{pmatrix}$ & $\frac{11}{4}$ & $-$20.67  & $-$17.60 \\
\hline\hline
\end{tabular}
\end{table}
To validate the reliability of the present approach, the spatial distribution of the spin density is visualized via the spin density isosurfaces shown in Fig.~\ref{Fig:spin_densities}.  The resulting spin density maps confirm that the spin density remains well localized on the respective Cu(${\mathrm{II}}$) ions throughout the self-consistent field (SCF) procedure.

Figure~\ref{Fig:Broken_sym_energies} shows the total energies of all six spin states relative to the BS$_4$ configuration. The resulting energy values by DFT are reported in Table~\ref{tab:relative_energies}. Across both exchange--correlation functionals, the HS configuration shows the highest relative total energy, while the BS$_5$ configuration corresponds to the lowest-energy state. Consequently, the system favors a low-spin configuration, yielding an $S = 1/2$ (doublet) ground state. This result is in agreement with the $\chi T(T)$ 
measurements on the \cu\ powder sample described in Sec.~\ref{sec: Static Magnetic Properties}.
It is notable that PBE0 yields a slightly larger $E_{HS}-E_{BS_5}$ energy separation (i.e., about 11 meV larger) due to its higher and distance independent fraction of exact exchange, which enhances spin localization and strengthens superexchange interactions compared to the range-separated CAM-B3LYP functional.

We have employed Eq.~\ref{eqn:J_AB} to determine the corresponding radial and tangential exchange coupling parameters. 
For the radial exchange coupling between the central  Cu$({\mathrm{II})}$ ion and the peripheral ions we substituted 
the corresponding total energies and expectation values of the HS and BS$_1$ spin states from Table~\ref{tab:relative_energies}. For the tangential  exchange coupling constant $J_2$ between the peripheral Cu(II) we  used the HS and BS$_4$ states. The choice of the BS1 state is made because it leads to spin-flipped configuration for all $J_1$ couplings. BS$_4$ is chosen because it has one spin flip in tangential direction for each spin at the corners. 
The resulting theoretical magnetic exchange values are reported in Table~\ref{tab:J_values}. 


\begin{table}[b]
    \centering
    \caption[]{Exchange coupling constants for the \cu\ obtained using  PBE0 and CAM-B3LYP hybrid functionals in comparison to values obtained from fitting the experimental $\chi T(T)$ data.}
    \begin{tabular*}{\columnwidth}{@{\extracolsep{\fill}}c|cc}   \hline\hline     
        Method & $J_1$ (meV) & $J_2$ (meV) \\ \hline
        PBE0 & $-$28.98 & $-$19.42 \\
        CAM-B3LYP & $-$24.56 & $-$16.50 \\ \hline
        $\chi T$ & $-$33.5(8) & $-$18.2(5) \\\hline\hline
    \end{tabular*}
    \label{tab:J_values}
\end{table}

With the choice of BS states above, this approach yields a relation of $J_1/J_2 \approx 1.49$ for both the PBE0 and CAM-B3LYP functionals, correctly reproducing the antiferromagnetic nature of both couplings. The obtained interaction constants are in good agreement with the experimental values yielding $J_1$/$J_2$ = 1.84(7) (see Tab.~\ref{tab:J_values}), thus corroborating the analysis of experimental data and validating our DFT calculations. 
Similar results have been reported in previous studies on the Cu(II)[12-MC\textsubscript{Cu(II)N(Shi)}-4]$^{2-}$ metallacrown complex~\cite{Happ2014}, where the calculated exchange parameters for the bulk state generally agreed with those obtained from $\chi T(T)$ measurements on powder (bulk) samples~\cite{Happ2015,Happ2016,Tavakoli}, with the remaining discrepancies between theory and experiment usually attributed to limitations of the DFT exchange-correlation functionals~\cite{Pavlyukh2019}.

\subsection{$g$-tensor of \cu}

The general form of the interaction of an electronic magnetic moment with an external magnetic field $\mathbf{B}$ is given by Zeeman Hamiltonian~\cite{Harriman1978}: 
\begin{equation}\label{eq:spin_Hamiltonian_external_field}
    \mathcal{H_Z} = \mu_B \mathbf{B} \cdot \mathbf{g} \cdot \mathbf{S}
\end{equation}
where $\mathbf{S}$ is the electronic spin, $\mu_B$ is the Bohr magneton, and $\mathbf{g}$ is the electronic $g$-tensor. 
The $g$-tensor for the coupling to an external magnetic field
that is determined in EPR spectroscopy can also be computed \textit{ab initio}~\cite{Schreckenbach}. The paramagnetic contribution to $g$-tensor, which arises from the combined influence of orbital Zeeman effect and spin-orbit coupling, can be expressed as a mixed second derivative with respect to the external magnetic field and the electron spin~\cite{Neese2001,Pavlyukh2019}.
\begin{equation}\label{eq:g_value}
    g = \frac{1}{\mu_B} \left( \frac{\partial^2 E}{\partial B \partial S} \right)_{B=0, S=0}
\end{equation}
This second-derivative expression is evaluated by the coupled perturbed spin-density self-consistent field (CP-SCF) approach introduced by Neese~\cite{Neese2001} for the calculation of EPR $g$-tensors and implemented in the ORCA code~\cite{ORCA}. We have used this code employing Kohn-Sham methods for the self-consistent calculations.

\begin{table}[b]
    \centering
    \caption[]{Calculated principal components of the EPR $g$-tensor and the corresponding isotropic $g$-factor ($g_{\mathrm{iso}}$) obtained using the PBE0 and CAM-B3LYP hybrid functionals}
    \begin{tabular*}{\columnwidth}{@{\extracolsep{\fill}}l|cccc}   \hline\hline     
        Method & $g_x$ & $g_y$ & $g_z$ & $g_{\mathrm{iso/eff}}$ \\ \hline
        PBE0 & 2.061 & 2.100 & 2.168 & 2.110 \\
        CAM-B3LYP & 2.060 & 2.098 & 2.164 & 2.107 \\ \hline\hline
    \end{tabular*}
    \label{tab:g_values}
\end{table}

The calculated principal components of the $g$-tensor obtained with the PBE0 and CAM-B3LYP functionals, as listed in Table~\ref{tab:g_values}, display moderate anisotropy, characterized by a pronounced enhancement of $g_z$ relative to $g_x$ and $g_y$. The deviation from the free-electron value indicates substantial spin-orbit coupling contributions, giving rise to a magnetic easy axis along the $z$ direction which is perpendicular to the \cu\ metallacrown plane. The $g$-tensor obtained from the DFT calculations in Table~\ref{tab:g_values} is in good agreement with the g-factors extracted from the ground-state EPR signal $s_1$ (see Tab.~\ref{tab:Cu5_fitparameter}) and reproduces the axial splitting, which is slightly stronger in the experimental $g$-factors. In particular, the effective isotropic $g$-factor $g_{\mathrm{iso/eff}}$ is in excellent agreement with the experimental result of $g = 2.109(13)$, derived from the static susceptibility data (See Sec.~\ref{sec:EPR-Spectroscopy}) and coinciding with EPR results.
The close correspondence between theory and experiment indicates that both hybrid functionals provide a reliable description of the magnetic response of the system.

\section{electronic structure}\label{sec:electronic structure}

In this section, we clarify the spin resolved electronic structure as it is obtained from the DFT calculations as implemented in the \textsc{ORCA} quantum chemistry program~\cite{ORCA}. We focus only on our results using PBE0 hybrid functional~\cite{PBE0}. 
The calculated results presented here provide a basis for interpreting future scanning tunneling spectroscopy and spin and time resolved photoemission experiments. Since experimental data are not yet available, in order to obtain a smooth representation of the calculated electronic spectrum, we applied Gaussian smearing with a full width at half maximum (FWHM) of 0.52~eV.

\begin{figure}[tb]
	\centering
	\includegraphics[width=1\columnwidth]{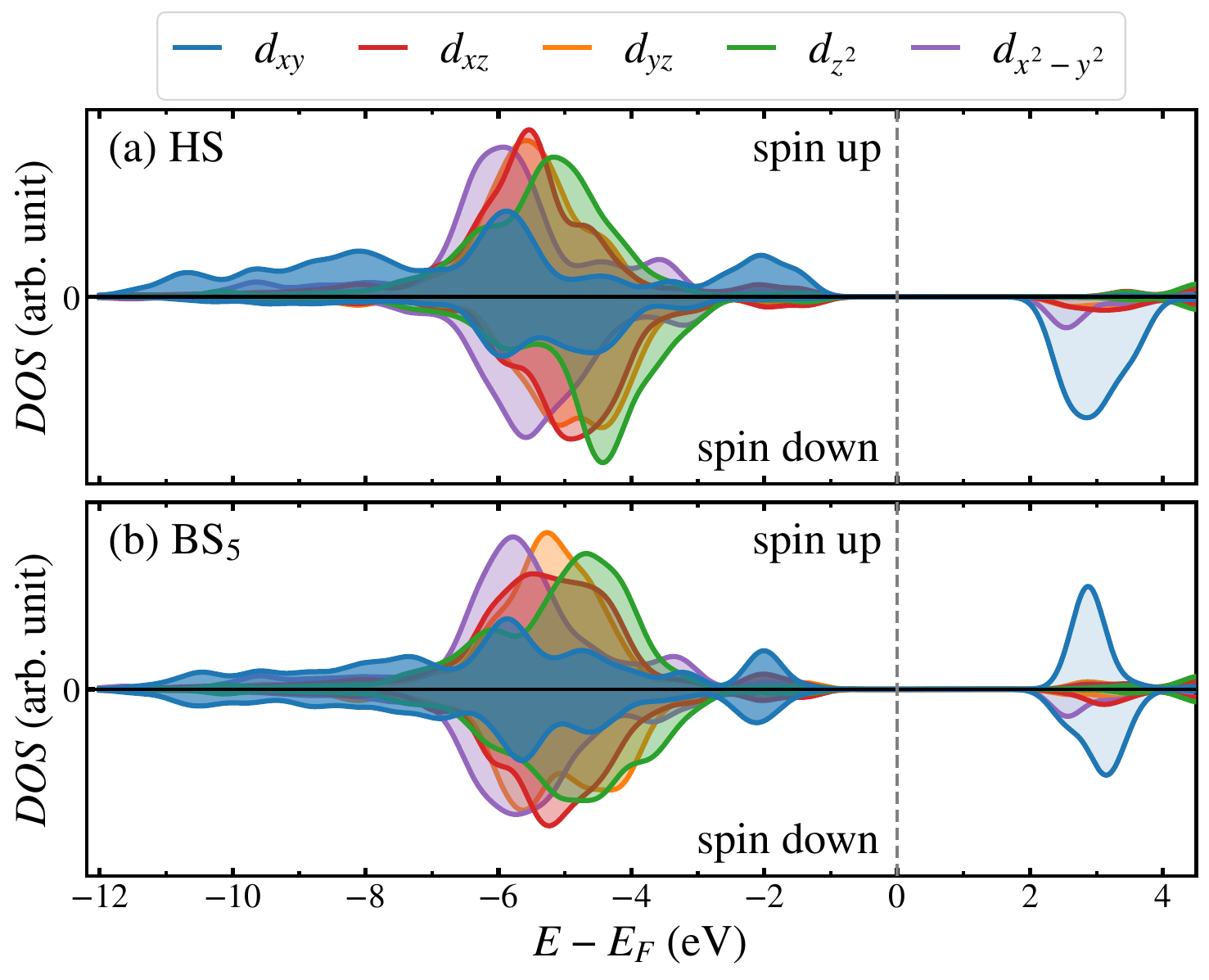}
	\caption{
    Theoretical spin-resolved projected density of states (pDOS) of the Cu $3d$ orbitals in Cu$_4$Cu, obtained from ORCA calculations using the PBE0 exchange correlation functional. (a) High-spin (HS) configuration with fully aligned  Cu$({\mathrm{II})}$ spins ($S = 5/2$) and (b) BS$_5$ configuration ($S = 1/2$ ground state), obtained by flipping the spins of the diagonal  Cu$({\mathrm{II})}$ ions in the peripheral ring (also see Table~\ref{tab:relative_energies}). 
    The vertical dashed line denotes the Fermi energy $E_{\rm F}$.
    }
\label{Fig:partial_d_BS5_HS}
\end{figure}

\begin{figure}[tb]
	\centering
	\includegraphics[width=1\columnwidth]{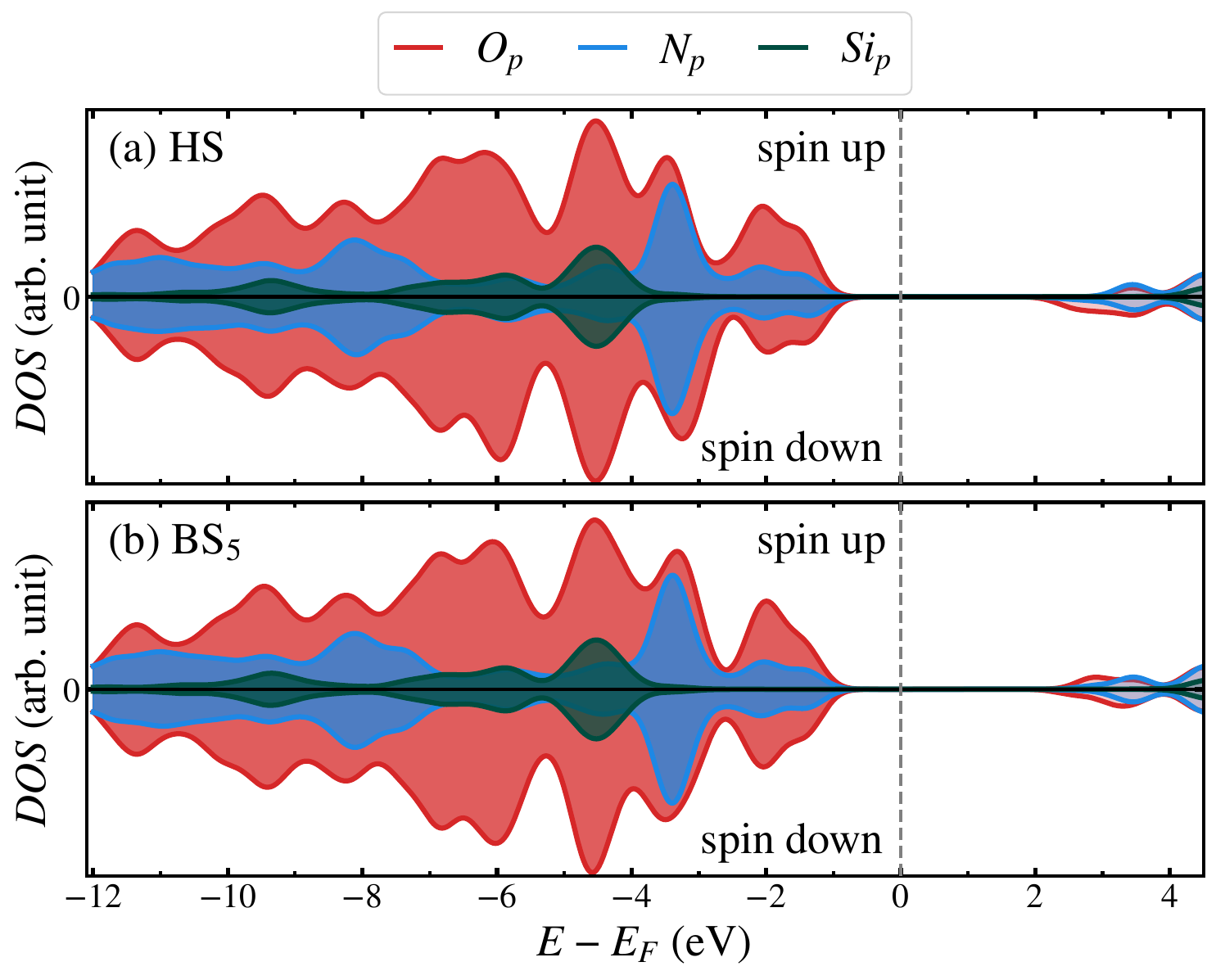}
	\caption{Theoretical spin-resolved pDOS of the ligand $p$-orbitals (O, N, and Si) in \cu\, calculated at the ORCA PBE0 level. Here,  (a) and (b) correspond to the same BS configurations as shown in Fig.~\ref{Fig:partial_d_BS5_HS}. 
    The vertical dashed line denotes the Fermi level. 
    }
\label{fig:P_orbitals_ONSi}
\end{figure}

Figures~\ref{Fig:partial_d_BS5_HS} and \ref{fig:P_orbitals_ONSi} show the spin-resolved density of states (DOS) of two representative spin states: the ferromagnetic high-spin (HS) state and the BS$_5$ configuration identified as the lowest-energy broken-symmetry state, as discussed in~\ref{sec:ground-state}.

Figure ~\ref{Fig:partial_d_BS5_HS} details the pDOS of $3d$-orbitals of the  Cu$({\mathrm{II})}$ ions. The HS configuration, for which the  Cu$({\mathrm{II})}$ spins are fully aligned ($S = 5/2$), exhibits a pronounced exchange splitting with the occupied spin-up $d$-states predominantly located in the energy window between approximately $-4$ and $-7$~eV, see Fig.~\ref{Fig:partial_d_BS5_HS}(a). Moreover, a strong spin-up polarization in the $d_{xy}$ orbital is observed at around $-2$~eV. In the unoccupied region above the Fermi level $E_{\text{F}}$, a distinct peak with strong $d_{xy}$ orbital character appears exclusively in the spin-down channel, indicating a well-defined spin polarization of this orbital in the HS configuration. 

The pDOS of the spin states BS$_5$ in Fig.~\ref{Fig:partial_d_BS5_HS}(b) with $S=1/2$ also shows occupied $d$-states in a valence energy range similar to that of the HS state, but exhibits noticeable modifications in the spin-dependent electronic structure especially at around $-2$~eV, where a characteristic energy splitting between both spin channels of $d_{xy}$ orbital is visible. Above $E_{\rm F}$, the BS$_5$ configuration exhibits clear peaks in both spin channels, too, indicating mainly $d_{xy}$-character and spin-splitting of the unoccupied orbitals.


Figure~\ref{fig:P_orbitals_ONSi} shows the spin-resolved pDOS of the ligand $p$-orbitals, i.e., O, N, and Si atoms.
In the HS state (Fig. ~\ref{fig:P_orbitals_ONSi}(a)), the valence region between approximately $-10$ and $-2$~eV is dominated by O$_p$ states, with noticeable but smaller contributions from N$_p$ orbitals and comparatively weaker participation of Si$_p$ states. The broad distribution of O$_p$ orbital contributions indicates significant Cu–O hybridization between the metal $d$-manifold and ligand $p$-orbitals. Nitrogen $p$-states contribute prominently in the energy window around $-4$ to $-3$~eV, overlapping with the Cu $d$-states and suggesting their involvement in magnetic exchange pathways. The Si$_p$ contributions are more localized, reflecting their more peripheral electronic role. Moreover, clear spin polarization of the ligand $p$-states is observed in the HS configuration, particularly for oxygen atoms.

In the $BS_5$ configuration shown in Fig.~\ref{fig:P_orbitals_ONSi}(b) the overall energetic distribution of the ligand $p$-states remains largely preserved, demonstrating that the fundamental bonding framework is maintained upon spin inversion of selected Cu centers. However, minor changes in the spin-resolved spectral density are evident. Especially, above the Fermi level ($E_{\rm F}$), only minor ligand $p$-orbital contributions are present in both configurations. In the O$_p$ orbitals, a persistent small spin polarization and spin splitting are observed, which indicates the exchange splitting due to the flipped spins of the peripheral Cu$({\mathrm{II}})$ ions, cf. Fig.~\ref{Fig:spin_densities}.

\section{Discussion and Conclusions}

\begin{figure}[t]
	\centering
	\includegraphics[width=1\columnwidth]{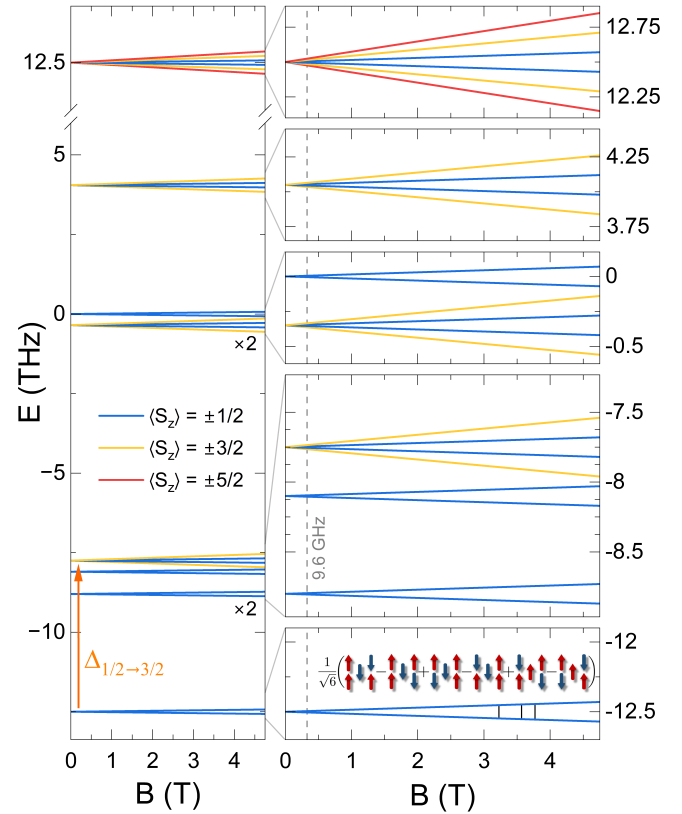}
	\caption{Energy-level diagram for \cu\ obtained by exact diagonalization of the spin Hamiltonian \eqref{eqn:Cu5_SH_SQUID}. 
    The left-hand side shows the full energy spectrum of \eqref{eqn:Cu5_SH_SQUID} with degeneracies denoted if present and the lowest-energy doublet-to-quartet transition of excitation energy $\Delta_{1/2\rightarrow3/2}$ indicated by a vertical arrow. The right-hand side panels zoom into the close-lying multiplets with the X-band EPR transition field for $f = 9.6$~GHz is marked by the vertical dashed lines.  In the lower-right panel showing the ground state doublet energy levels, the transitions observed in HF-EPR are represented by black vertical lines, and the wave function of one of the ground states is depicted.}
	\label{Fig:Cu5_eleveldiagram}
\end{figure}

The spin Hamiltonian Eq. \eqref{eqn:Cu5_SH_SQUID} for \cu\ and the magnetic couplings $J_i$ -- obtained experimentally and validated by our numerical study -- provide a coherent explanation of the  observations in magnetometry and EPR spectroscopy. So far, the electronic structure was discussed in terms of the broken-symmetry (BS) states obtained from DFT calculations (Sec.~\ref{ch:numerics}). These BS states are not eigenstates of the total spin and contain some effects of strong correlations among the electrons. Using exact diagonalization of the spin Hamiltonian \eqref{eqn:Cu5_SH_SQUID}, we now proceed with a discussion of our Heisenberg model of \cu\ in an external magnetic field $\mathbf{B}$. The states discussed in the following are, thus, eigenstates of the total spin and do not exhibit a broken symmetry.
Using the exchange couplings $J_1 = -8.1(2)$~THz ($-33.5(8)$~meV), $J_2 = -4.40(12)$~THz ($-18.2(5)$~meV), and the effective isotropic $g$-factor, $g=2.109$, from the analysis of the static susceptibility data (Fig.~\ref{Fig:Cu5_dc_susceptibility}) yields the energy spectrum of \eqref{eqn:Cu5_SH_SQUID} consisting of $32=2^5$ eigenstates, which is shown in Fig.~\ref{Fig:Cu5_eleveldiagram}.


The ground state in zero magnetic field is a spin doublet at $E/h=-12.5$~THz; its wavefunction is a superposition of the four states with the central Cu$({\mathrm{II})}$ spin and one outer Cu$({\mathrm{II})}$ spin flipped (as BS\textsubscript{3}) and the two states with opposite Cu$({\mathrm{II})}$ spins flipped (as BS\textsubscript{5}), as sketched in Fig.~\ref{Fig:Cu5_eleveldiagram}. Toward higher energy, the ground state is followed by two doublets (one of them degenerate) and a quartet, and further multiplets at even higher energies.
The thermal population of this first quartet over the energy gap $\Delta_{1/2\rightarrow3/2}/h = 4.8(4)~\text{THz}$ ($\Delta/k_{\rm B}=230(19)$~K) from the ground state doublet leads to the increase in $\chi T$ starting above 30~K. The highest multiplet of the spin system is at 25~THz ($1200$~K) above the ground state doublet, which agrees with the non-saturating behavior of  $\chi T$ up to room temperature.

With increasing magnetic field, the multiplets split in energy. The low-temperature signal $s_1$ observed in X-band EPR and HF-EPR (Figs.~\ref{Fig:Cu5_EPR_all_spectra} and \ref{Fig:Cu5_HFEPR}) originates from the spin transition within the ground state doublet. The thermally-induced signal $s_2$ observed above 33~K can be attributed to one (or several) transitions within the multiplets between $-$9~THz and $-$7.5~THz, which are separated not more than $\Delta_{1/2\rightarrow3/2}$ from the ground state doublet. 
As anisotropic $g$-tensors are not included in the spin Hamiltonian \eqref{eqn:Cu5_SH_SQUID} but replaced by an effective $g$-factor identical for all five Cu$({\mathrm{II})}$ spins, the observed anisotropy of EPR transitions $s_1$ and $s_2$ is not present in Fig.~\ref{Fig:Cu5_eleveldiagram}. Advancing to a more sophisticated spin model for \cu, one may include individual $g$-tensors for each  Cu$({\mathrm{II})}$ to explain the anisotropic and different $g$-factors of ground state ($s_1$) and excited state transitions ($s_2$) reported herein, and could proceed to exchange interactions beyond isotropic coupling.

In conclusion, we report the synthesis, the structural characterization, as well as dynamic and static magnetic properties of the metallacrown Cu$({\mathrm{II})}$[12-MC\textsubscript{Cu$({\mathrm{II})}$N(L)}-4]$^{2-}$ (\cu), which provides an excellent platform to investigate finite clusters of paramagnetic \cu\ centers in the solid state, as intermolecular magnetic interactions are inhibited by bulky ligands.
Static magnetization data reveal competing antiferromagnetic exchange interaction between the five Cu(II) ions of $J_1=-8.1(2)$~THz 
($-33.5(8)$~meV) 
and $J_2=-4.40(12)$~THz 
($-18.2(5)$~meV) 
inducing a doublet ground state in \cu. Using X-band and high-frequency/high-field electron paramagnetic resonance, we determine the $g$-factors $g_x = 2.028$, $g_y = 2.056$, and $g_z = 2.222$ of the ground-state, which are close to axial anisotropy. Our \textit{ab initio} broken symmetry-based DFT calculations corroborate the antiferromagnetic nature of magnetic interactions with coinciding $J_1/J_2$ ratio, confirm the $S=1/2$ ground state, and yield $g$-factors in agreement with the experimental results. 
EPR spectroscopy further reveals a thermally excited signal above 33~K, which is associated with transitions to excited doublet and quartet states at
around 4.8(4)~\text{THz} 
above the ground state.
We moreover numerically study the spin resolved electronic structure which in particular elucidates the orbital character of the unoccupied (magnetic) orbitals and may be used in future experimental investigations. Overall, this work yields an experimentally and numerically validated quantitative model for the ground state and excited state magnetism in the \cu\ metallacrown, thereby providing a full picture on magnetism in this paradigmatic Cu(II) cluster with competing exchange interactions.

\section*{Acknowledgements}
Support by the Deutsche Forschungsgemeinschaft (DFG) under Germany's Excellence Strategy EXC2181/1-390900948 (The Heidelberg STRUCTURES Excellence Cluster) and the TRR 173–268565370 Spin + X (projects A08, A09, MGK) is gratefully acknowledged. All DFT calculations were performed on the high-performance computing cluster Elwetritsch at RPTU Kaiserslautern-Landau. A CPU grant for the project 'Spin+SMM' through the Alliance for High Performance Computing Rheinland-Pfalz (AHRP) is gratefully acknowledged. A.T. thanks P. Puschnig (Graz) for hospitality and helpful discussions. L.B. and R.K. acknowledge support by the DFG via the Research Training Group GRK 2948/1 'Mixed Ionic Electronic Transport'.

\bibliography{metallacrown}

@article{Tegoni2012,
title = {Metallacrowns of copper(II) and aminohydroxamates: Thermodynamics of self assembly and host–guest equilibria},
journal = {Coordination Chemistry Reviews},
volume = {256},
number = {1},
pages = {289-315},
year = {2012},
issn = {0010-8545},
doi = {https://doi.org/10.1016/j.ccr.2011.06.007},
url = {https://www.sciencedirect.com/science/article/pii/S001085451100169X},
author = {Matteo Tegoni and Maurizio Remelli}
}

@article{Bodwin2001,
title = {The development of chiral metallacrowns into anion recognition agents and porous materials},
journal = {Coordination Chemistry Reviews},
volume = {216-217},
pages = {489-512},
year = {2001},
issn = {0010-8545},
doi = {https://doi.org/10.1016/S0010-8545(00)00396-9},
url = {https://www.sciencedirect.com/science/article/pii/S0010854500003969},
author = {Jeffrey J Bodwin and Annabel D Cutland and Roneil G Malkani and Vincent L Pecoraro}
}

@Article{Puetz2025,
AUTHOR = {Pütz, Frederik and Blättner, Richard and Kurek, Yves and Bolz, Lukas and Ehnert, Swen and Wendels, Robert and Stephan, Dominic and Schreyer, Philip and Ranecki, Robert and Brennfleck, Ellen and Lüpke, Anne and Laible, Dominik and Baumann, Benedikt and Lach, Stefan and Rentschler, Eva and Ziegler, Christiane},
TITLE = {Liquid Phase Preparation of Organic Thin Films Consisting of Complex Molecules—The Example of the Metallacrown {CuCu\textsubscript{4}}},
JOURNAL = {Solids},
VOLUME = {6},
YEAR = {2025},
NUMBER = {1},
pages = {13},
URL = {https://www.mdpi.com/2673-6497/6/1/13},
ISSN = {2673-6497},
DOI = {10.3390/solids6010013}
}

@article{Ranecki2023,
    author = {Ranecki, Robert and Lach, Stefan and Lüpke, Anne and Rentschler, Eva and Ziegler, Christiane},
    title = {Spin-Flip Inelastic Electron Tunneling Spectroscopy on a {CuCu\textsubscript{4}} Metallacrown Complex on {Au(111)}},
    journal = {The Journal of Physical Chemistry C},
    volume = {127},
    number = {27},
    pages = {13186-13195},
    year = {2023},
    month = {07},
    issn = {1932-7447},
    doi = {10.1021/acs.jpcc.3c01291},
    url = {https://doi.org/10.1021/acs.jpcc.3c01291},
    eprint = {https://pubs.acs.org/jpccck/article-pdf/127/27/13186/2879770/jp3c01291.pdf},
}

@article{Perdew_2021,
   title={Interpretations of ground-state symmetry breaking and strong correlation in wavefunction and density functional theories},
   volume={118},
   ISSN={1091-6490},
   url={http://dx.doi.org/10.1073/pnas.2017850118},
   DOI={10.1073/pnas.2017850118},
   pages = {e2017850118},
   number={4},
   journal={Proceedings of the National Academy of Sciences},
   publisher={National Academy of Sciences},
   author={Perdew, John P. and Ruzsinszky, Adrienn and Sun, Jianwei and Nepal, Niraj K. and Kaplan, Aaron D.},
   year={2021},
   month=Jan }

@article{Happ2016,
  title = {Analyzing the enforcement of a high-spin ground state for a metallacrown single-molecule magnet},
  author = {Happ, P. and Sapozhnik, A. and Klanke, J. and Czaja, P. and Chernenkaya, A. and Medjanik, K. and Schuppler, S. and Nagel, P. and Merz, M. and Rentschler, E. and Elmers, H. J.},
  journal = {Phys. Rev. B},
  volume = {93},
  issue = {17},
  pages = {174404},
  numpages = {7},
  year = {2016},
  month = {May},
  publisher = {American Physical Society},
  doi = {10.1103/PhysRevB.93.174404},
  url = {https://link.aps.org/doi/10.1103/PhysRevB.93.174404}
}

@Article{Happ2014,
author ="Happ, Peter and Rentschler, Eva",
title  ="Enforcement of a high-spin ground state for the first 3d heterometallic 12-metallacrown-4 complex",
journal  ="Dalton Trans.",
year  ="2014",
volume  ="43",
issue  ="41",
pages  ="15308-15312",
publisher  ="The Royal Society of Chemistry",
doi  ="10.1039/C4DT02275K",
url  ="http://dx.doi.org/10.1039/C4DT02275K"}

@article{Neese2001,
    author = {Neese, Frank},
    title = {Prediction of electron paramagnetic resonance g values using coupled perturbed Hartree–Fock and Kohn–Sham theory},
    journal = {The Journal of Chemical Physics},
    volume = {115},
    number = {24},
    pages = {11080-11096},
    year = {2001},
    month = {12},
    issn = {0021-9606},
    doi = {10.1063/1.1419058},
    url = {https://doi.org/10.1063/1.1419058}
}

@misc{SM,
  author       = "",
  title        = "",
  howpublished = "",
  month        = "",
  year         = "",
  note         = "See supplemental material at [url will be inserted by publisher] which contains additional information on the synthesis, the structure as well as further numerical data."
}

@article{Schreckenbach,
author = {Schreckenbach, Georg and Ziegler, Tom},
title = {Calculation of the G-Tensor of Electron Paramagnetic Resonance Spectroscopy Using Gauge-Including Atomic Orbitals and Density Functional Theory},
journal = {The Journal of Physical Chemistry A},
volume = {101},
number = {18},
pages = {3388-3399},
year = {1997},
doi = {10.1021/jp963060t}
}

@book{Harriman1978,
  title = {Theoretical Foundations of Electron Spin Resonance},
  author = {Harriman, John E.},
  year = {1978},
  publisher = {Academic Press},
  address = {New York}
}

@book{goodenough,
  title={Magnetism and the Chemical Bond},
  author={Goodenough, John B},
  year={1963},
  publisher={Interscience Publishers},
  address={New York},
  series={Interscience Monographs on Chemistry},
  volume={1}
}

@article{Ju2024,
  title = {{EPR} spectroscopy: A versatile tool for exploring transition metal complexes in organometallic and bioinorganic chemistry},
  volume = {45},
  ISSN = {1229-5949},
  url = {http://dx.doi.org/10.1002/bkcs.12899},
  DOI = {10.1002/bkcs.12899},
  number = {10},
  journal = {Bulletin of the Korean Chemical Society},
  publisher = {Wiley},
  author = {Ju,  Minyoung and Kim,  Jin and Shin,  Jeongcheol},
  year = {2024},
  pages = {835–862}
}

@article{CAM-B3LYP,
title = {A new hybrid exchange–correlation functional using the Coulomb-attenuating method {(CAM-B3LYP)}},
journal = {Chemical Physics Letters},
volume = {393},
number = {1},
pages = {51-57},
year = {2004},
issn = {0009-2614},
doi = {https://doi.org/10.1016/j.cplett.2004.06.011},
url = {https://www.sciencedirect.com/science/article/pii/S0009261404008620},
author = {Takeshi Yanai and David P Tew and Nicholas C Handy}
}

@article{PBE0,
    author = {Adamo, Carlo and Barone, Vincenzo},
    title = {Toward reliable density functional methods without adjustable parameters: The {PBE0} model},
    journal = {The Journal of Chemical Physics},
    volume = {110},
    number = {13},
    pages = {6158-6170},
    year = {1999},
    month = {04},
    issn = {0021-9606},
    doi = {10.1063/1.478522},
    url = {https://doi.org/10.1063/1.478522}
}

@article{ORCA,
author = {Neese, Frank},
title = {The {ORCA} program system},
journal = {WIREs Computational Molecular Science},
volume = {2},
number = {1},
pages = {73-78},
doi = {https://doi.org/10.1002/wcms.81},
url = {https://wires.onlinelibrary.wiley.com/doi/abs/10.1002/wcms.81},
eprint = {https://wires.onlinelibrary.wiley.com/doi/pdf/10.1002/wcms.81},
year = {2012}
}

@incollection{Yamaguchi2018,
author = {Yasutaka Kitagawa and Toru Saito and Kizashi Yamaguchi},
title = {Approximate Spin Projection for Broken-Symmetry Method and Its Application},
booktitle = {Symmetry (Group Theory) and Mathematical Treatment in Chemistry},
publisher = {IntechOpen},
address = {London},
year = {2018},
editor = {Takashiro Akitsu},
chapter = {7},
doi = {10.5772/intechopen.75726},
url = {https://doi.org/10.5772/intechopen.75726}
}

@article{Pavlyukh2019,
  title = {Broken symmetry states of metallacrowns: Distribution of spins and the $g$ tensor},
  author = {Pavlyukh, Y. and Rentschler, E. and Elmers, H. J. and H\"ubner, W. and Lefkidis, G.},
  journal = {Phys. Rev. B},
  volume = {99},
  issue = {14},
  pages = {144418},
  numpages = {12},
  year = {2019},
  month = {Apr},
  publisher = {American Physical Society},
  doi = {10.1103/PhysRevB.99.144418},
  url = {https://link.aps.org/doi/10.1103/PhysRevB.99.144418}
}

@article{WHANGBO2003,
title = {Spin exchange interactions and magnetic structures of extended magnetic solids with localized spins: theoretical descriptions on formal, quantitative and qualitative levels},
journal = {Journal of Solid State Chemistry},
volume = {176},
number = {2},
pages = {417-481},
year = {2003},
issn = {0022-4596},
doi = {https://doi.org/10.1016/S0022-4596(03)00273-1},
url = {https://www.sciencedirect.com/science/article/pii/S0022459603002731},
author = {Myung-Hwan Whangbo and Hyun-Joo Koo and Dadi Dai}
}

@article{Dai2003,
    author = {Dai, Dadi and Whangbo, Myung-Hwan},
    title = {Spin exchange interactions of a spin dimer: Analysis of broken-symmetry spin states in terms of the eigenstates of Heisenberg and Ising spin Hamiltonians},
    journal = {The Journal of Chemical Physics},
    volume = {118},
    number = {1},
    pages = {29-39},
    year = {2003},
    month = {01},
    issn = {0021-9606},
    doi = {10.1063/1.1525809},
    url = {https://doi.org/10.1063/1.1525809}
}

@article{Noodleman1986,
title = {Ligand spin polarization and antiferromagnetic coupling in transition metal dimers},
journal = {Chemical Physics},
volume = {109},
number = {1},
pages = {131-143},
year = {1986},
issn = {0301-0104},
doi = {https://doi.org/10.1016/0301-0104(86)80192-6},
url = {https://www.sciencedirect.com/science/article/pii/0301010486801926},
author = {Louis Noodleman and Ernest R. Davidson}
}

@article{Noodleman1981,
    author = {Noodleman, Louis},
    title = {Valence bond description of antiferromagnetic coupling in transition metal dimers},
    journal = {The Journal of Chemical Physics},
    volume = {74},
    number = {10},
    pages = {5737-5743},
    year = {1981},
    month = {05},
    issn = {0021-9606},
    doi = {10.1063/1.440939},
    url = {https://doi.org/10.1063/1.440939}
}

@article{Tavakoli,
author = {Tavakoli, Ariyan and Lach, Stefan and Pütz, Frederik and Stephan, Dominic and Lüpke, Anne and Laible, Dominik and Kaidisch, Siegfried and Windischbacher, Andreas and Rentschler, Eva and Puschnig, Peter and Schneider, Hans-Christian and Ziegler, Christiane},
title = {Investigating the Structural Factors Influencing the Magnetic Interaction between Individual Copper Spins in the {CuCu\textsubscript{4}} Metallacrown Complex, from Its Bulk Form to Its Adsorption on {Au(111)}},
journal = {Advanced Materials Interfaces},
volume = {13},
number = {4},
pages = {e00906},
doi = {https://doi.org/10.1002/admi.202500906},
url = {https://advanced.onlinelibrary.wiley.com/doi/abs/10.1002/admi.202500906},
eprint = {https://advanced.onlinelibrary.wiley.com/doi/pdf/10.1002/admi.202500906},
year = {2026}
}

@article{Sayre1955,
author = {Sayre, Ralph},
title = {The Identity of Heilpern's “Pinacolylthiourea” and the Preparation of Authentic 2-Thiono-4,4,5,5-tetramethylimidazolidine},
journal = {Journal of the American Chemical Society},
volume = {77},
number = {24},
pages = {6689-6690},
year = {1955},
doi = {10.1021/ja01629a112},
URL = {https://doi.org/10.1021/ja01629a112}
}

@article{Ullman1972,
author = {Ullman, Edwin F. and Osiecki, Jeanne H. and Boocock, D. G. B. and Darcy, R.},
title = {Stable free radicals. {X}. Nitronyl nitroxide monoradicals and biradicals as possible small molecule spin labels},
journal = {Journal of the American Chemical Society},
volume = {94},
number = {20},
pages = {7049-7059},
year = {1972},
doi = {10.1021/ja00775a031},
URL = {https://doi.org/10.1021/ja00775a031}
}

@article{Kolanji2017,
author = {Kolanji, Kubandiran and Ravat, Prince and Bogomyakov, Artem S. and Ovcharenko, Victor I. and Schollmeyer, Dieter and Baumgarten, Martin},
title = {Mixed Phenyl and Thiophene Oligomers for Bridging Nitronyl Nitroxides},
journal = {The Journal of Organic Chemistry},
volume = {82},
number = {15},
pages = {7764-7773},
year = {2017},
doi = {10.1021/acs.joc.7b00435},
URL = {https://doi.org/10.1021/acs.joc.7b00435}
}

@article{Hirel2001,
author = {Hirel, Catherine and Vostrikova, Kira E. and Pécaut, Jacques and Ovcharenko, Victor I. and Rey, Paul},
title = {Nitronyl and Imino Nitroxides: Improvement of Ullman's Procedure and Report on a New Efficient Synthetic Route},
journal = {Chemistry – A European Journal},
volume = {7},
number = {9},
pages = {2007-2014},
doi = {10.1002/1521-3765(20010504)7:93.0.CO;2-7},
year = {2001}
}

@article{vanderSluis1990,
  title = {BYPASS: an effective method for the refinement of crystal structures containing disordered solvent regions},
  volume = {46},
  ISSN = {0108-7673},
  url = {http://dx.doi.org/10.1107/S0108767389011189},
  DOI = {10.1107/s0108767389011189},
  number = {3},
  journal = {Acta Crystallographica Section A Foundations of Crystallography},
  publisher = {International Union of Crystallography (IUCr)},
  author = {van der Sluis,  P. and Spek,  A. L.},
  year = {1990},
  month = mar,
  pages = {194–201}
}

@article{Dolomanov2009,
  title = {OLEX2: a complete structure solution,  refinement and analysis program},
  volume = {42},
  ISSN = {0021-8898},
  url = {http://dx.doi.org/10.1107/S0021889808042726},
  DOI = {10.1107/s0021889808042726},
  number = {2},
  journal = {Journal of Applied Crystallography},
  publisher = {International Union of Crystallography (IUCr)},
  author = {Dolomanov,  Oleg V. and Bourhis,  Luc J. and Gildea,  Richard J. and Howard,  Judith A. K. and Puschmann,  Horst},
  year = {2009},
  month = jan,
  pages = {339–341}
}

@article{werner2017,
  title = {{Anisotropy-governed competition of magnetic phases in the honeycomb quantum magnet ${\mathrm{Na}}_{3}{\mathrm{Ni}}_{2}{\mathrm{SbO}}_{6}$ studied by dilatometry and high-frequency ESR}},
  author = {Werner, J. and Hergett, W. and Gertig, M. and Park, J. and Koo, C. and Klingeler, R.},
  journal = {Phys. Rev. B},
  volume = {95},
  issue = {21},
  pages = {214414},
  numpages = {7},
  year = {2017},
  month = {Jun},
  publisher = {American Physical Society},
  doi = {10.1103/PhysRevB.95.214414}
}

@article{Sheldrick2015b,
  title = {Crystal structure refinement with SHELXL},
  volume = {71},
  ISSN = {2053-2296},
  url = {http://dx.doi.org/10.1107/S2053229614024218},
  DOI = {10.1107/s2053229614024218},
  number = {1},
  journal = {Acta Crystallographica Section C Structural Chemistry},
  publisher = {International Union of Crystallography (IUCr)},
  author = {Sheldrick,  George M.},
  year = {2015},
  month = jan,
  pages = {3–8}
}

@article{Wang2018,
  title = {Magnetoelastic coupling and ferromagnetic-type in-gap spin excitations in multiferroic $\alpha$-{Cu$_2$V$_2$O$_7$}},
  volume = {20},
  ISSN = {1367-2630},
  url = {http://dx.doi.org/10.1088/1367-2630/aac9dc},
  DOI = {10.1088/1367-2630/aac9dc},
  number = {6},
  journal = {New Journal of Physics},
  publisher = {IOP Publishing},
  author = {Wang,  L and Werner,  J and Ottmann,  A and Weis,  R and Abdel-Hafiez,  M and Sannigrahi,  J and Majumdar,  S and Koo,  C and Klingeler,  R},
  year = {2018},
  month = jun,
  pages = {063045}
}

@article{Sheldrick2015a,
  title = {SHELXT– Integrated space-group and crystal-structure determination},
  volume = {71},
  ISSN = {2053-2733},
  url = {http://dx.doi.org/10.1107/S2053273314026370},
  DOI = {10.1107/s2053273314026370},
  number = {1},
  journal = {Acta Crystallographica Section A Foundations and Advances},
  publisher = {International Union of Crystallography (IUCr)},
  author = {Sheldrick,  George M.},
  year = {2015},
  month = jan,
  pages = {3–8}
}

@article{Fidan2016,
  title = {Synthesis and Straightforward Quantification Methods of Imino Nitroxide-Based Hexaradical Architecture on a Cyclotriphosphazene Scaffold},
  volume = {55},
  ISSN = {1520-510X},
  url = {http://dx.doi.org/10.1021/acs.inorgchem.6b01976},
  DOI = {10.1021/acs.inorgchem.6b01976},
  number = {21},
  journal = {Inorganic Chemistry},
  publisher = {American Chemical Society (ACS)},
  author = {Fidan,  Ismail and \"{O}nal,  Emel and Yerli,  Yusuf and Luneau,  Dominique and Ahsen,  Vefa and Hirel,  Catherine},
  year = {2016},
  month = oct,
  pages = {11447–11453}
}

@article{Osiecki1968,
author = {Osiecki, Jeanne H. and Ullman, Edwin F.},
title = {Studies of free radicals. {I}. $\alpha$-Nitronyl nitroxides, a new class of stable radicals},
journal = {Journal of the American Chemical Society},
volume = {90},
number = {4},
pages = {1078-1079},
year = {1968},
doi = {10.1021/ja01006a053},
URL = { https://doi.org/10.1021/ja01006a053}
}

@article{Happ2015,
title = {{12-MC-4} metallacrowns as versatile tools for {SMM} research},
journal = {Coordination Chemistry Reviews},
volume = {289-290},
pages = {238-260},
year = {2015},
issn = {0010-8545},
doi = {https://doi.org/10.1016/j.ccr.2014.11.012},
url = {https://www.sciencedirect.com/science/article/pii/S0010854514003324},
author = {Peter Happ and Christian Plenk and Eva Rentschler}
}

@article{Koo2016,
author = {Koo ,Changhyun and Zvereva ,Elena A. and Shukaev ,Igor L. and Richter ,Michael and Stratan ,Mikhail I. and Vasiliev ,Alexander N. and Nalbandyan ,Vladimir B. and Klingeler ,R\"{u}diger},
title = {Static and Dynamic Magnetic Response of Fragmented Haldane-like Spin Chains in Layered {Li\textsubscript{3}Cu\textsubscript{2}SbO\textsubscript{6}}},
journal = {Journal of the Physical Society of Japan},
volume = {85},
number = {8},
pages = {084702},
year = {2016},
doi = {10.7566/JPSJ.85.084702},
URL = {      https://doi.org/10.7566/JPSJ.85.084702}
}

@article{Koo2019,
title = {High-frequency EPR study on {Cu\textsubscript{4}Cu}- and {Co\textsubscript{4}Co}-metallacrown complexes},
journal = {Journal of Magnetism and Magnetic Materials},
volume = {477},
pages = {340-343},
year = {2019},
issn = {0304-8853},
doi = {https://doi.org/10.1016/j.jmmm.2019.01.038},
url = {https://www.sciencedirect.com/science/article/pii/S0304885318324065},
author = {C. Koo and J. Park and J. Butscher and E. Rentschler and R. Klingeler}
}

@article{stoll_easyspin_2006,
	title = {{EasySpin}, a comprehensive software package for spectral simulation and analysis in {EPR}},
	volume = {178},
	issn = {10907807},
	doi = {10.1016/j.jmr.2005.08.013},
	pages = {42--55},
	number = {1},
	journal = {Journal of Magnetic Resonance},
	shortjournal = {Journal of Magnetic Resonance},
	author = {Stoll, Stefan and Schweiger, Arthur},
	urldate = {2021-03-23},
	year = {2006},
	langid = {english}
}

@article{bain_diamagnetic_2008,
	title = {Diamagnetic Corrections and Pascal's Constants},
	volume = {85},
	issn = {0021-9584, 1938-1328},
	url = {https://pubs.acs.org/doi/abs/10.1021/ed085p532},
	doi = {10.1021/ed085p532},
	pages = {532},
	number = {4},
	journal = {Journal of Chemical Education},
	shortjournal = {J. Chem. Educ.},
	author = {Bain, Gordon A. and Berry, John F.},
	urldate = {2021-09-06},
	date = {2008-04},
	langid = {english}
}

@article{mezei_structural_2007,
	title = {Structural and Functional Evolution of Metallacrowns},
	volume = {107},
	issn = {0009-2665, 1520-6890},
	url = {https://pubs.acs.org/doi/10.1021/cr078200h},
	doi = {10.1021/cr078200h},
	pages = {4933--5003},
	number = {11},
	journal = {Chemical Reviews},
	shortjournal = {Chem. Rev.},
	author = {Mezei, Gellert and Zaleski, Curtis M. and Pecoraro, Vincent L.},
	urldate = {2021-10-01},
	date = {2007-11-01},
    year = {2007}
}

@article{lah_isolation_1989,
	title = {Isolation and characterization of \{{Mn\textsuperscript{II}}[{Mn\textsuperscript{III}} (salicylhydroximate)]\textsubscript{4}(acetate) \textsubscript{2}({DMF})\textsubscript{6}\}$\cdot$2{DMF}: an inorganic analog of {M}\textsuperscript{2+}(12-crown-4)},
	volume = {111},
	issn = {0002-7863, 1520-5126},
	url = {https://pubs.acs.org/doi/abs/10.1021/ja00200a054},
	doi = {10.1021/ja00200a054},
	shorttitle = {Isolation and characterization of \{{MnII}[{MnIII}(salicylhydroximate)]4(acetate)2({DMF})6\}.cntdot.2DMF},
	pages = {7258--7259},
	number = {18},
	journal = {Journal of the American Chemical Society},
	shortjournal = {J. Am. Chem. Soc.},
	author = {Lah, Myoung Soo and Pecoraro, Vincent L.},
	urldate = {2021-10-01},
	year = {1989}
}

@article{zaleski_synthesis_2004,
	title = {Synthesis, Structure, and Magnetic Properties of a Large Lanthanide–Transition-Metal Single-Molecule Magnet},
	volume = {43},
	issn = {1433-7851, 1521-3773},
	url = {https://onlinelibrary.wiley.com/doi/10.1002/anie.200454013},
	doi = {10.1002/anie.200454013},
	pages = {3912--3914},
	number = {30},
	journal = {Angewandte Chemie International Edition},
	shortjournal = {Angew. Chem. Int. Ed.},
	author = {Zaleski, Curtis M. and Depperman, Ezra C. and Kampf, Jeff W. and Kirk, Martin L. and Pecoraro, Vincent L.},
	urldate = {2021-10-01},
	year = {2004},
	langid = {english}
}

@article{boron_mixed_2010,
	title = {A Mixed 3d-4f 14-Metallacrown-5 Complex That Displays Slow Magnetic Relaxation through Geometric Control of Magnetoanisotropy},
	volume = {49},
	issn = {0020-1669, 1520-510X},
	url = {https://pubs.acs.org/doi/10.1021/ic101121d},
	doi = {10.1021/ic101121d},
	pages = {9104--9106},
	number = {20},
	journal = {Inorganic Chemistry},
	shortjournal = {Inorg. Chem.},
	author = {Boron, Thaddeus T. and Kampf, Jeff W. and Pecoraro, Vincent L.},
	urldate = {2021-10-01},
	year = {2010},
	langid = {english}
}

@article{deb_understanding_2014,
	title = {Understanding Spin Structure in Metallacrown Single-Molecule Magnets using Magnetic Compton Scattering},
	volume = {136},
	issn = {0002-7863, 1520-5126},
	url = {https://pubs.acs.org/doi/10.1021/ja501452w},
	doi = {10.1021/ja501452w},
	pages = {4889--4892},
	number = {13},
	journal = {Journal of the American Chemical Society},
	shortjournal = {J. Am. Chem. Soc.},
	author = {Deb, Aniruddha and Boron, Thaddeus T. and Itou, Masayoshi and Sakurai, Yoshiharu and Mallah, Talal and Pecoraro, Vincent L. and Penner-Hahn, James E.},
	urldate = {2021-10-01},
	year = {2014},
	langid = {english}
}

@article{wang_opening_2021,
	title = {Opening Magnetic Hysteresis by Axial Ferromagnetic Coupling: From Mono‐Decker to Double‐Decker Metallacrown},
	volume = {60},
	issn = {1433-7851, 1521-3773},
	url = {https://onlinelibrary.wiley.com/doi/10.1002/anie.202014993},
	doi = {10.1002/anie.202014993},
	shorttitle = {Opening Magnetic Hysteresis by Axial Ferromagnetic Coupling},
	pages = {5299--5306},
	number = {10},
	journal = {Angewandte Chemie International Edition},
	shortjournal = {Angew. Chem. Int. Ed.},
	author = {Wang, Jin and Li, Quan‐Wen and Wu, Si‐Guo and Chen, Yan‐Cong and Wan, Rui‐Chen and Huang, Guo‐Zhang and Liu, Yang and Liu, Jun‐Liang and Reta, Daniel and Giansiracusa, Marcus J. and Wang, Zhen‐Xing and Chilton, Nicholas F. and Tong, Ming‐Liang},
	urldate = {2021-10-01},
	year = {2021},
	langid = {english}
}

@article{song_preparation_2000,
	title = {Preparation, crystal structures and magnetic properties of 12-metallacrown-4 complexes with the donors on the organic periphery of molecule},
	volume = {305},
	issn = {00201693},
	url = {https://linkinghub.elsevier.com/retrieve/pii/S0020169300001249},
    year= {2000},
	doi = {10.1016/S0020-1693(00)00124-9},
	pages = {135--142},
	number = {2},
	journal = {Inorganica Chimica Acta},
	shortjournal = {Inorganica Chimica Acta},
	author = {Song, You and Liu, Jia-Cheng and Liu, Yong-Jiang and Zhu, Dun-Ru and Zhuang, Jin-Zhong and You, Xiao-Zeng},
	urldate = {2021-10-05},
	date = {2000-07},
	langid = {english}
}

@article{pavlishchuk_magnetic_2011,
	title = {Magnetic and Sorption Properties of Supramolecular Systems Based on Pentanuclear Copper({II}) 12-Metallacrown-4 Complexes and Isomeric Phthalates: Structural Modeling of the Different Stages of Alcohol Sorption},
	year = {2011},
	issn = {14341948},
	url = {https://onlinelibrary.wiley.com/doi/10.1002/ejic.201100790},
	doi = {10.1002/ejic.201100790},
	shorttitle = {Magnetic and Sorption Properties of Supramolecular Systems Based on Pentanuclear Copper({II}) 12-Metallacrown-4 Complexes and Isomeric Phthalates},
	pages = {4826--4836},
	number = {31},
	journal = {European Journal of Inorganic Chemistry},
	shortjournal = {Eur. J. Inorg. Chem.},
	author = {Pavlishchuk, Anna V. and Kolotilov, Sergey V. and Zeller, Matthias and Shvets, Oleksiy V. and Fritsky, Igor O. and Lofland, Samuel E. and Addison, Anthony W. and Hunter, Allen D.}
}

@article{mcdonald_progressive_2013,
	title = {Progressive decoration of pentanuclear {Cu(II)} 12-metallacrown-4 nodes towards targeted 1- and 2D extended networks},
	volume = {15},
	issn = {1466-8033},
	url = {http://xlink.rsc.org/?DOI=c3ce40859k},
	doi = {10.1039/c3ce40859k},
	pages = {6672},
	number = {34},
	journal = {{CrystEngComm}},
	shortjournal = {{CrystEngComm}},
	author = {{McDonald}, Cecelia and Whyte, Teresa and Taylor, Stephanie M. and Sanz, Sergio and Brechin, Euan K. and Gaynor, Declan and Jones, Leigh F.},
	urldate = {2021-10-05},
	year = {2013},
	langid = {english}
}

@article{liu_crystal_2020,
	title = {Crystal structure and magnetic properties of a pentanuclear copper({II}) 12-metallacrown-4 complex derived from glutaroimide-dioxime},
	volume = {118},
	issn = {13877003},
	url = {https://linkinghub.elsevier.com/retrieve/pii/S138770032030575X},
	doi = {10.1016/j.inoche.2020.107985},
	pages = {107985},
	journal = {Inorganic Chemistry Communications},
	shortjournal = {Inorganic Chemistry Communications},
	author = {Liu, Xiao-Meng and Li, Xue-Qi and Wang, Qing-Lun and Liu, Qian-Ying and Tong, Yu-Zhang and Jia, Xiao-Pu and Yang, Chun},
	urldate = {2021-10-05},
	year = {2020},
	langid = {english}
}

@Article{Comba_HFESR_2015,
  author    = {Peter Comba and Michael Gro{\ss}hauser and Rüdiger Klingeler and Changhyun Koo and Yanhua Lan and Dennis Müller and Jaena Park and Annie Powell and Mark J. Riley and Hubert Wadepohl},
  journal   = {Inorg. Chem.},
  title     = {Magnetic Interactions in a Series of Homodinuclear Lanthanide Complexes},
  year      = {2015},
  month     = nov,
  number    = {23},
  pages     = {11247--11258},
  volume    = {54},
  doi       = {10.1021/acs.inorgchem.5b01673},
  publisher = {American Chemical Society ({ACS})},
}

\hfill
\newpage

\end{document}